\documentclass[twocolumn,aps, reprint ,superscriptaddress,floatfix]{revtex4-1} 
\usepackage{amsmath}
\usepackage{amssymb}
\usepackage{comment}
\usepackage{multirow}% multi row text in tables
\usepackage{lipsum}% just for random text  
\usepackage{braket}% dirac notation
\usepackage{xcolor}% Include comments with different color
\usepackage{graphicx}% Include figure files
\usepackage{dcolumn}% Align table columns on decimal point
\usepackage{bm}% bold math
\usepackage[colorlinks = true, ,citecolor = blue, linkcolor = blue, urlcolor = blue]{hyperref}% add hypertext capabilities
\begin{document}

%\title{Topological Phase Transition in Non-Hermitian Pseudospin-Polarized Waveguides}% Force line breaks with \\

\title{Non-Hermitian Photonic Spin Hall Insulators}

%\title{Non-Hermitian $\mathbb{Z}_2$ Photonic Topological Insulators}% Force line breaks with \\

\author{Rodrigo P. Câmara}
\affiliation{Instituto Superior Técnico and Instituto de Telecomunicações, University of Lisbon, Avenida Rovisco Pais 1, Lisboa, 1049001 Portugal}

\author{Tatiana G. Rappoport}
\affiliation{Centro de Física das Universidade do Minho e do Porto (CFUMUP) e Departamento de Física, Universidade do Minho, P-4710-057 Braga, Portugal}
\affiliation{Instituto de F\'\i sica, Universidade Federal do Rio de Janeiro, C.P. 68528, 21941-972 Rio de Janeiro RJ, Brazil}

\author{M\'ario G. Silveirinha}
\affiliation{Instituto Superior Técnico and Instituto de Telecomunicações, University of Lisbon, Avenida Rovisco Pais 1, Lisboa, 1049001 Portugal}
	
\date{\today}% It is always \today, today,
             %  but any date may be explicitly specified

\begin{abstract}
Photonic platforms invariant under parity ($\mathcal{P}$), time-reversal ($\mathcal{T}$), and duality ($\mathcal{D}$) can support topological phases analogous to those found in time-reversal invariant ${\mathbb{Z}_2}$ electronic systems with conserved spin. Here, we demonstrate the resilience of the underlying spin Chern phases against non-Hermitian effects, notably material dissipation. We identify that non-Hermitian, $\mathcal{P}\mathcal{D}$-symmetric, and reciprocal photonic insulators fall into two topologically distinct classes. Our analysis focuses on the topology of a $\mathcal{P}\mathcal{D}$-symmetric and reciprocal parallel-plate waveguide (PPW). We discover a critical loss level in the plates that marks a topological phase transition. The Hamiltonian of the $\mathcal{P}\mathcal{T}\mathcal{D}$-symmetric system is found to consist of an infinite direct sum of Kane-Mele type Hamiltonians with a common band gap. This structure leads to the topological charge of the waveguide being an ill-defined sum of integers due to the particle-hole symmetry. Each component of this series corresponds to a spin-polarized edge state. Our findings present a unique instance of a topological photonic system that can host an infinite number of edge states in its band gap.
\end{abstract}

%\begin{abstract}
%Photonic platforms invariant under parity ($\mathcal{P}$), time-reversal ($\mathcal{T}$), and duality ($\mathcal{D}$) can support topological phases analogous to those found in time-reversal invariant ${\mathbb{Z}_2}$ electronic systems. Here, we demonstrate the resilience of ${\mathbb{Z}_2}$ phases against non-Hermitian effects, notably material dissipation. We identify that non-Hermitian $\mathcal{P}\mathcal{D}$-symmetric and reciprocal photonic insulators fall into two topologically distinct classes, with interfaces capable of supporting spin-polarized states. Our analysis focuses on the topology of a $\mathcal{P}\mathcal{D}$-symmetric parallel-plate waveguide (PPW). We discover a critical loss level in the plates that marks a transition between different topological phases. The Hamiltonian of the system is found to consist of an infinite direct sum of Kane-Mele type Hamiltonians with a common band gap. This structure leads to the topological charge of a ($\mathcal{P}\mathcal{T}\mathcal{D}$-symmetric) waveguide being an ill-defined sum of integers due to the particle-hole symmetry. Each component of this series corresponds to a spin-polarized edge state. Our findings present a unique instance of a topological photonic system that can host an infinite number of edge states in its band gap.
%\end{abstract}

%\keywords{Suggested keywords}%Use showkeys class option if keyword
                              %display desired
\maketitle

%%%%%%%%%%%%%%%%%%%%%%%%%%%%%%%%%%%%%%%%%%%%%%%%%%%%%%%%%%%%%%%%%
% INTRODUCTION
%%%%%%%%%%%%%%%%%%%%%%%%%%%%%%%%%%%%%%%%%%%%%%%%%%%%%%%%%%%%%%%%%

Topological photonics \cite{review_Ling_Lu, review_Kim} establishes a unique paradigm to create unidirectional channels immune to back-scattering. The topological protection of Chern insulators originates from an electromagnetic ``quantum Hall effect'' rooted in the time-reversal symmetry breaking \cite{Haldane_crystals, Raghu_crystals}. 
Time-reversal invariant photonic structures also may host a plethora of topological phases \cite{reciprocal_2, reciprocal_3, insulators_1, insulators_2, insulators_3, insulators_4, insulators_5, insulators_11, insulators_7, insulators_8, Silveirinha_continuous_Z2}, such as the optical counterpart of the quantum spin Hall state \cite{kane_mele_z2_qsh, QSHE_Kane_Mele}.
%However, 
Due to the distinct nature of fermionic and bosonic systems, additional symmetries, such as duality symmetry which ensures balanced electric and magnetic responses, are essential for establishing ${\mathbb{Z}_2}$ topological protection in photonic platforms \cite{Silveirinha_ptd, insulators_3}.
 
%In general, time-reversal invariant electronic systems are characterized by a $\mathbb{Z}_2$-valued topological index. 

To our best knowledge, the most general symmetry transformation that can enable
%enables
the ${\mathbb{Z}_2}$ topological protection in photonic systems is
%determined by 
a combination of the parity ($\mathcal{P}$), time-reversal ($\mathcal{T}$), and duality ($\mathcal{D}$) operators, $\tilde{\mathcal{T}} = \mathcal{P} \mathcal{T} \mathcal{D}$ \cite{Silveirinha_ptd, Chen2015, insulators_3, sylvain_kane_mele}. The $\tilde {\cal T}$ operator may be regarded as a pseudo-time-reversal operator, as it is anti-linear and satisfies ${{\tilde {\cal T}}^2} = - {\bf{1}}$.
This property ensures that photonic states in a ${\mathcal{P}}{\mathcal{T}}{\mathcal{D}}$ symmetric  system are degenerate, in accordance with Kramers' theorem \cite{Silveirinha_ptd, insulators_3}. 
Moreover, it implies the existence of a basis where the system's scattering matrix is anti-symmetric \cite{Silveirinha_ptd}. Importantly, this characteristic indicates that in systems with an odd number of bi-directional propagation channels, light transport can occur without back-reflections \cite{Silveirinha_ptd}.

Recently, the study of topological phases was extended to non-Hermitian systems through a generalization of the notion of band gaps to complex spectra \cite{Silveirinha_non_hermitian, non_hermitian_theory, non_hermitian_comment, selective_enhancement_1, selective_enhancement_2, non_hermitian_comment, exotic_EP_1_top, exotic_EP_3_top, non_hermitian_theory, non_hermitian_poss, top_open_1, top_open_2, top_open_3, top_open_5, fractional_charge}. However, previous studies on $\mathcal{P}\mathcal{T}\mathcal{D}$-symmetric systems dealt exclusively with energy conserving (Hermitian) platforms \cite{energy_sinks, surface_impedances, dual_surfaces, Chen2015, sylvain_kane_mele}. 
%As a result, topological properties with no counterpart in nonreciprocal \cite{Haldane_crystals, Raghu_crystals, nonreciprocal_1, nonreciprocal_2, nonreciprocal_3, nonreciprocal_4, nonreciprocal_5, Silveirinha_bulk_edge} and time-reversal invariant \cite{reciprocal_2, reciprocal_3, insulators_1, insulators_2, insulators_3, insulators_4, insulators_5, insulators_6, insulators_7, insulators_8} closed systems have been found \cite{non_hermitian_comment, selective_enhancement_1, selective_enhancement_2, non_hermitian_comment, exotic_EP_1_top, exotic_EP_3_top, non_hermitian_theory, non_hermitian_poss, top_open_1, top_open_2, top_open_3, top_open_5, fractional_charge}. 
%Various photonic ${\mathbb{Z}_2}$ topological insulators have been explored \cite{energy_sinks, surface_impedances, dual_surfaces, Chen2015, sylvain_kane_mele}. 
Here, we study the impact of non-Hermitian effects, notably material dissipation, on the topological phases of $\mathcal{P}\mathcal{T}\mathcal{D}$-symmetric systems.
In time-reversal invariant electronic systems, the topological protection of the edge channels is stronger when %if 
spin is also conserved \cite{ct_chan_2022}. A mapping can be established between the $\mathbb{Z}_2$ invariant and the $\mathbb{Z}$-valued Chern numbers associated with the spin sectors \cite{z2_spin_chern_number}.
%, but the latter ultimately determine the edge phenomena. 
The topological indices of the spin sectors offer the most comprehensive description of the properties of the edge states.
In $\mathcal{P}\mathcal{T}\mathcal{D}$-invariant optical platforms, light modes also have a preserved polarization (pseudospin). Therefore, we focus on the $\mathbb{Z}$-valued Chern numbers of the pseudospin sectors. We discover that these 
%${\mathbb{Z}_2}$ topological index
topological indices demonstrate remarkable resilience to material absorption. Notably, we identify a topological phase transition controlled by the strength of the loss parameter. This transition is characterized by exceptional mode degeneracies and resonant energy absorption.
Moreover, the gap ``charge'' of the topological phase is characterized by a non-convergent sum of integers. 
This distinct feature is a consequence of the particle-hole symmetry of the spectrum of photonic systems 
and is manifested by the emergence of an infinite number of edge states within the topological band gap.

%Our focus is on modifying the structure introduced in Ref. \onlinecite{Chen2015}, adding dissipation. Spin-polarized unidirectional edge modes in $\mathcal{P}\mathcal{T}\mathcal{D}$-symmetric PPWs have been studied previously \cite{Chen2015, Silveirinha_exact_edge_PTD}, yet a proper topological classification remains elusive.
%even in the Hermitian case.
%In this work, we begin by noting that the electromagnetic fields are also spin-polarized in the more general $\mathcal{P}\mathcal{D}$-symmetric PPWs. We classify the topology by determining the band and gap Chern numbers of the dispersion diagram, for the two spin sectors. There are two topological phases that depend on the loss parameter and are separated by a transition tied to exceptional degeneracies and resonant energy absorption. In particular, we find that the phase underlying the lossless ($\mathcal{P}\mathcal{T}\mathcal{D}$-symmetric) guide is robust to dissipation. Even though its gap Chern number is an ill-defined infinite sum of integers, we show that the edge transport along an anisotropic and $\mathcal{P}\mathcal{T}\mathcal{D}$-symmetric wall boundary is in agreement with the bulk-edge correspondence.

\begin{figure}[t]
\includegraphics[width=1.\columnwidth,clip]{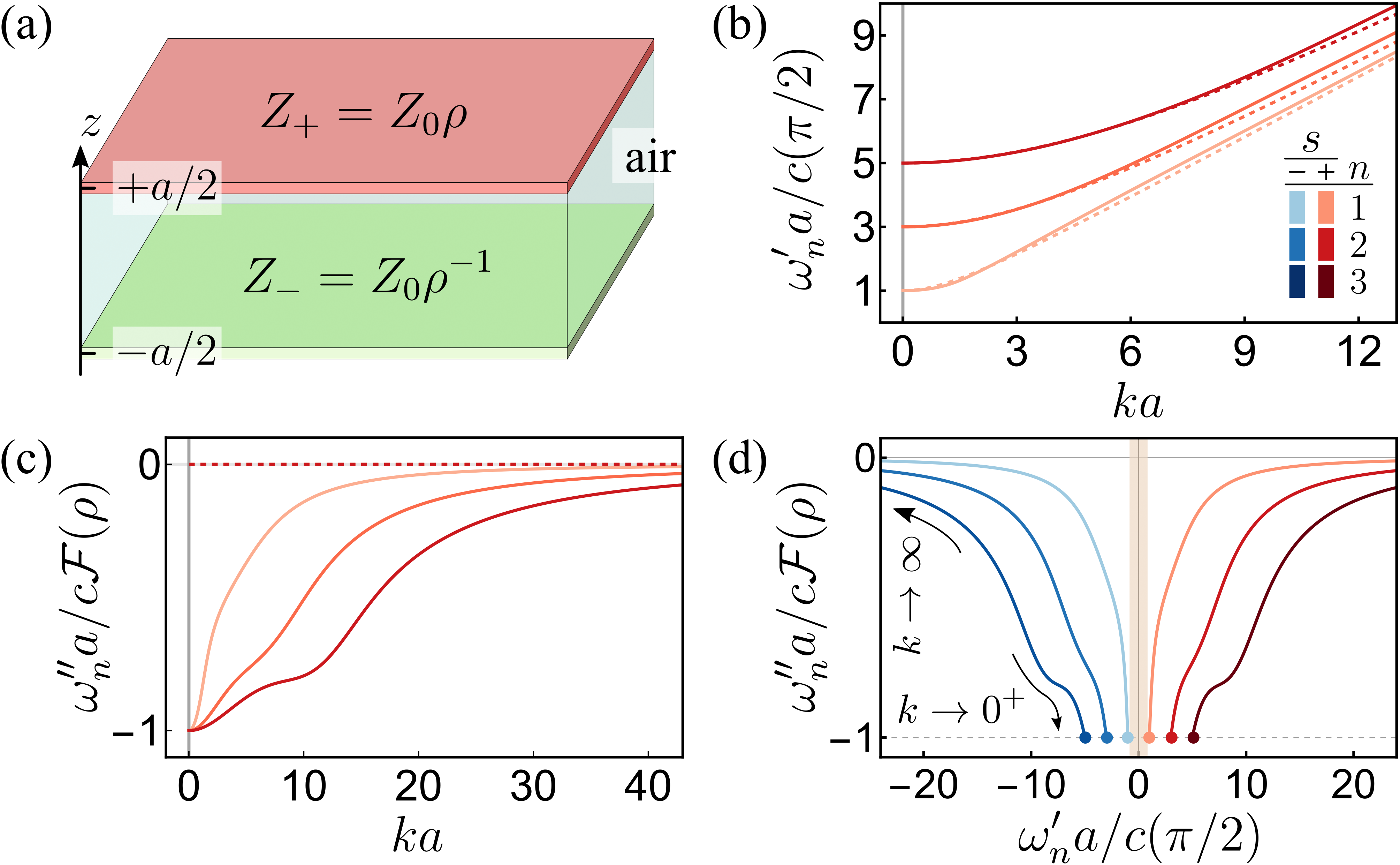}
\caption{\label{fig::band.structure} (a) Structure of the ${\cal P}{\cal D}$-symmetric parallel-plate wave\-guide with a dielectric block (air) of width $a$ in between two impedance surfaces. (b) Real $\omega'_n$ and (c) imaginary $\omega''_n$ parts of the eigenfrequencies as functions of $k$. We choose $s = +$ (positive bands) and consider the first few guided modes: $n=1,2,3$. Dashed (Solid) curves refer to $\rho = 0$ ($\rho = 0.6$). $\omega''_n$ in (c) is normalized to the function $\mathcal{F}(\rho)$. (d) Projected band structure in the complex plane for the case $\rho = 0.6$. Positive (Negative) frequency branches are in red (blue). The frequency values for $k=0$ are marked by solid dots.}
\end{figure}

 Our system consists of a parallel-plate waveguide [see Fig. \hyperref[fig::band.structure]{1(a)}], where the electromagnetic responses of the top and bottom plates are interconnected through electromagnetic duality \cite{Chen2015, Silveirinha_exact_edge_PTD}. Each plate is modeled by an impedance boundary condition, defined as $Z_i \, \bm{\hat{n}} \times \bm{H} = \bm{E}_{\tan}$ with $\bm{\hat{n}}$ the unit normal vector to the plate, oriented  towards the dielectric. %$\bm{E}_{\tan} = \bm{E} - \bm{\hat{ n}} \left(\bm{\hat{n}} \cdot \bm{E} \right)$ 
 Here, $\bm{E}_{\tan}$ is the tangential electric field, and $Z_{i}$ the surface impedance of the $i= +/-$ (top/bottom, respectively) plate. Similarly to Ref. \cite{Silveirinha_ptd}, we consider the parity transformation $\left( {x,y,z} \right) \to \left( {x,y, - z} \right)$ that exchanges the positions of the plates, and the duality mapping
${\cal D}:\left( \bm{E},\bm{H} \right) \to \left( \bm{H} Z_0, - \bm{E}/ Z_0 \right)$ that converts ${Z_+} \to {Z_-}$ and vice-versa ($Z_0$ is the vacuum impedance). The composition of these two transformations leaves the air region of the guide unaltered. The system is parity-duality ($\mathcal{P}\mathcal{D}$) symmetric if and only if the two surface impedances satisfy $Z_{+} Z_{-} = Z_0^2$  \cite{SM}.
%This implies that the impedances of the plates are of the type $Z_{+} = Z_0\rho e^{i\phi}$ and $Z_{-} = Z_0\rho^{- 1} e^{- i\phi}$, with $\rho$ and $\phi$ dimensionless parameters. However, 

For simplicity, we assume the surface impedances $Z_{\pm}$ to be real and frequency-independent, defined as $Z_{\pm} = Z_0\rho^{\pm 1}$, 
where
%$Z_0$ is the vacuum impedance and 
$\rho$ is the normalized resistivity parameter.
When $\rho = 0$, the plates act as perfect electric (PEC) or magnetic (PMC) conductors, as in Ref. \cite{Chen2015}. 
In our analysis, $\rho$ is allowed to span real values. 
Positive $\rho$ values indicate that the plates absorb fields within the dielectric. 
Conversely, negative $\rho$ values correspond to active (amplifying) plates 
%In both cases, energy is either lost or gained, %breaking time-reversal symmetry. 
The waveguide is Hermitian and time-reversal invariant ($\mathcal{P}\mathcal{T}\mathcal{D}$-symmetric) only when $\rho = 0$. 
%We refer to $\rho$ as the guide's resistivity parameter henceforward.

In the following, we denote a point in space $\left( {x,y,z} \right)$ by $\left( {z} \right)$, and the mirror-symmetric point $\left( {x,y, - z} \right)$ by $\left( {-z} \right)$. For $\mathcal{P}\mathcal{D}$-symmetric and \emph{reciprocal} systems, such that 
$\varepsilon(z) = \mu(-z)$, the frequency-domain (source-free) Maxwell equations can be split into two independent sets of equations \cite{Chen2015}
\begin{equation}
\pm \frac{i c}{\varepsilon (z)}
\begin{pmatrix}
0 & \partial_z & \partial_y \\
- \partial_z & 0 & - \partial_x \\
\partial_y & - \partial_x & 0
\end{pmatrix} \bm{\Psi}^\pm(-z) = \omega \bm{\Psi}^\pm(z),
\label{eq::pseudospin_polarization}
\end{equation}

\noindent that are formally equivalent to $\hat{\mathcal{H}}^\pm \bm{\Psi}^\pm (z) \equiv \omega \bm{\Psi}^\pm(z)$ with the eigenstates $\bm{\Psi}^\pm$ written in terms of the electric and magnetic fields $\bm{E}$ and $\bm{H}$ as $\bm{\Psi}^\pm (z) = \left( E_x (z) \mp Z_0 H_x (-z), E_y (z) \mp Z_0 H_y (-z), \right.$ $\left.E_z (z) \pm Z_0 H_z (-z) \right)^\intercal$ \cite{Chen2015}. The ``polarization'' associated with the superscript ``$+$" or ``$-$" is conserved and represents an internal degree of freedom. We refer to it as pseudospin and to the modes $\bm{\Psi}^\pm$ as pseudospinors, following the terminology of topological photonics. Here, $c$ denotes the speed of light in vacuum and $\omega$ is the oscillation frequency. As $\bm{E} = \frac{1}{2}\left( \bm{\Psi}^{+} + \bm{\Psi}^{-} \right)$, the dynamics of the electric field is controlled by the dynamics of the two pseudospinors [Eq. (\ref{eq::pseudospin_polarization})]. Different from the original Maxwell's equations, the dynamics of the pseudospinors is strongly nonlocal, as the two sides of Eq. (\ref{eq::pseudospin_polarization}) are evaluated at mirror-symmetric points. Interestingly, the pseudospin decomposition remains valid even when the plate walls or the dielectric are lossy. 
%(e.g., if $\varepsilon$ is frequency dependent and complex-valued). 
In our guide, the dielectric is air ($\varepsilon = \mu = 1$), so non-Hermitian effects arise exclusively due to the plates. 

\begin{comment}
Suppose that the Chern numbers $\mathcal{C}^{\pm}$ associated with a frequency band of the isospectral operators $\mathcal{H}^{\pm}$ are well-defined. The electromagnetic reciprocity of the waveguide guarantees that $\mathcal{C}^{+} + \mathcal{C}^{-} = 0$, whether the electrodynamics is conservative ($\rho = 0$) or not ($\rho \neq 0$) \cite{Silveirinha_non_hermitian}. Althought the total Chern number vanishes, it is still possible for the band to be topologically nontrivial, i.e., characterized by $\mathcal{C}^{+} \neq 0$. In the conservative case, a $\mathcal{P}\mathcal{T}\mathcal{D}$-transformation switches the pseudospin of the eigenstates of the $\mathcal{P}\mathcal{T}\mathcal{D}$-symmetric guide. This is in analogy with the quantum spin Hall state of electronic systems \cite{QSHE_Kane_Mele, kane_mele_z2_qsh}. The state is time-reversal invariant and so the quantum Hall conductance is zero. However, spin is conserved and exchanged under time-reversal. Therefore, the spin Hall conductance is nonvanishing and quantized. From this analogy, we withdraw that the invariant $\nu \in \mathbb{Z}_2$ of the frequency band of the $\mathcal{P}\mathcal{T}\mathcal{D}$-symmetric guide relates to the Chern number of ``$+$" sector as $\nu = \mathcal{C}^{+} \, \text{mod} 2$ \cite{z2_spin_chern_number}. In the following, we apply non-Hermitian topological band theory to determine the topological phase diagram of $\mathcal{C}^{+}$ \cite{non_hermitian_theory}. 
\end{comment}

In the Hermitian case, where both the dielectric and the plates are lossless, the
%class of 
eigenstates with pseudospin ``+'' can be transformed into eigenstates with pseudospin ``-'' using the time-reversal operator. While such a construction is not feasible in the non-Hermitian case, a crucial observation is that the electromagnetic reciprocity of the system guarantees that the total topological charge vanishes \cite{Silveirinha_non_hermitian}, similar to the fermionic case \cite{spin_Chern_Haldane}. Thereby the Chern indices of the operators $\hat{\mathcal{H}}^\pm$ in Eq. (\ref{eq::pseudospin_polarization}) must be exactly balanced: ${\cal C}_{}^{\rm{ + }} + {\cal C}_{}^ -  = 0$. 
%Here, ${\cal C}_{}^ \pm $ are the Chern numbers associated with a band gap of interest. 
Therefore, ${\cal P}{\cal D}$-symmetric reciprocal systems may host nontrivial topological phases determined by the invariant ${\cal C}_{}^{\rm{ + }} =  - {\cal C}_{}^ - $. In the following, we apply the non-Hermitian topological band theory to determine the phase diagram of the nonlocal operators  $\hat{\mathcal{H}}^\pm$ \cite{non_hermitian_theory}. 

Since the system is invariant under continuous translations in the $xoy$ plane, the eigenstates can be factorized as $\bm{\Psi}^{\pm} (\bm{r},t) = \bm{\Psi}_{\bm{k}}^{\pm} (z) e^{-i \omega t} e^{i \bm{k} \cdot \bm{r}}$, where $\bm{k}$ is an in-plane real-valued wave vector of magnitude $k$. The pseudospinors are found by first solving Maxwell's equations subject to the appropriate boundary conditions in the guide, and then projecting the field solutions onto the pseudospinor subspaces \cite{SM}.
Because the PPW is also invariant under arbitrary rotations around the $z$-axis, it is convenient to write the eigenstate in terms of the unit vectors $(\bm{\hat{k}}, \bm{\hat{z}} \times \bm{\hat{k}}, \bm{\hat{z}})$. In this basis, the coordinates of the pseudospinors are
\begin{equation}
\begin{split}
    &\bm{\Psi}_{\bm{k}, n}^\pm (z)  \propto
    \begin{pmatrix}
    \mp \left( e^{-i \kappa_n z} + h_n e^{i \kappa_n z} \right) \\
    - \frac{\omega_n}{c \kappa_n} \left( e^{i \kappa_n z} - h_n e^{-i \kappa_n z}  \right) \\
    \mp \frac{k}{\kappa_n} \left( e^{-i \kappa_n z} - h_n e^{i \kappa_n z} \right)
    \end{pmatrix},
\end{split}
    \label{eq::pseudospin_envelopes}
\end{equation}
\noindent where $n = 1, 2, \dots$ labels different modes, $\kappa_n$ is a transverse wavenumber, $\omega_n$ is the eigenfrequency, and $h_n \equiv e^{i\kappa_n a} \frac{\omega_n - c \rho \kappa_n}{\omega_n + c \rho \kappa_{n}}$. For each $k$ real-valued, the transverse wavenumber satisfies the modal equation
\begin{equation}
    e^{2 i \kappa_n a} = \frac{\omega_n + c\rho \kappa_{n}}{\omega_n - c\rho \kappa_{n}} \frac{\rho \omega_n + c\kappa_{n}}{\rho\omega_n - c\kappa_{n}},
    \label{eq::modal_equation}
\end{equation}
\noindent with $\omega_n = s \times c \sqrt{k^2 + \kappa_{n}^2}$ and $s=\pm$. Each eigenmode is associated with an in-plane wave vector $\bm{k}$, with an integer $n$ that identifies the band, and with $s=\pm$ that specifies the square root branch. 
Importantly, the modal equation is independent of the pseudospin, resulting in an identical spectrum for the operators 
$\hat{\mathcal{H}}^\pm$.
%We see from the modal constraint that both $\kappa_n$ and $\omega_n$ are functions of the wavenumber $k$ and of the resistivity $\rho$. 
%Generally, it is impossible to express this dependence in closed-form because Eq. (\ref{eq::modal_equation}) is almost always transcendental. 
When $k = 0$, the modal equation yields analytical solutions $\kappa_n = (2n-1)\pi/2a - s \times i \mathcal{F}(\rho) /a$ \cite{SM} with 
%$s=\pm$ as before 
$\mathcal{F}(\rho) = \text{sgn}(\rho)
\ln \left| \frac{|\rho| + 1}{|\rho| -1} \right|$. These solutions are extended numerically to $k>0$ via a Nelder Mead minimization scheme \cite{SM}, allowing us to obtain the corresponding frequencies. We shall see that the singularities $\rho = \pm 1$ in the function $\mathcal{F} (\rho)$ play an important role in the topological properties of the waveguide.

Figures \hyperref[fig::band.structure]{1(b)} and \hyperref[fig::band.structure]{1(c)} show how the real $\omega_n'$ and imaginary $\omega_n''$ frequency parts vary with $k$, in systems with $\rho = 0$ (dashed lines) and $\rho = 0.6$ (solid lines). If the system is conservative ($\rho = 0$), the modal equation (\ref{eq::modal_equation}) reduces to $e^{2i\kappa_n a} = -1$, so the frequencies are real-valued, as expected.
%Moreover, they are odd multiples of $(c/a)(\pi/2)$ when $k = 0$, and the dispersion is that of the dielectric in the short wavelength limit ($\omega_n \sim s \times c k$ when $k \to \infty$). 
In the dissipative case ($\rho = 0.6$), the frequencies exhibit complex-values. %However, the dispersion of the dielectric is also recovered when $k \to \infty$. This is because the plates cannot dissipate the energy of the bulk waves when their wavelength is significantly shorter than the guide width $a$.

%To distinguish between the square root branches and avoid clumsy notation, we refer to the frequencies simply as positive/negative when $s = +/-$. 
Figure \hyperref[fig::band.structure]{1(d)} displays the projected band structure for $\rho = 0.6$, representing the locus of $\omega_n (k)$ for all $k$ real-valued. Notably, the diagram exhibits a mirror symmetry about the imaginary frequency axis: $\omega_n \to - \omega_n^*$. 
This symmetry, stemming from the reality of the electromagnetic field, correlates the positive and negative parts of the photonic spectrum ($s=\pm$), and is known as the particle-hole symmetry \cite{SM}. 
The positive and negative frequencies are separated by a gap on the complex plane (beige vertical strip). 
The frequency branches are disjoint, meaning they exhibit no intersections or self-intersections. These characteristics are typical of the band structure, except when the resistivity approaches $\rho = \pm 1$. 
%Nevertheless, the $n=1$ bands intersect when $\rho \sim \pm 1$, and the gap centered about the imaginary axis closes. We will explain this in detail further below and discuss how the nontrivial topology of the waveguide is affected. 
The band structure resides in the lower (upper) half of the complex plane when 
$\rho > 0$ ($\rho < 0$), indicative of the plates being lossy (gainy).

The topological classification of a generic non-Hermitian operator $\hat{\mathcal{H}} \neq \hat{\mathcal{H}}^\dagger$ requires a biorthogonal basis of left $\phi^\text{L}_n$ and right $\phi^\text{R}_n$ eigenstates, such that $\hat{\mathcal{H}}^\dagger \phi^\text{L}_n = E_n^* \phi^\text{L}_n$ and $\hat{\mathcal{H}} \phi^\text{R}_n = E_n \phi^\text{R}_n$, with $E_n$ generally complex-valued \cite{non_hermitian_theory}. In the supplementary material \cite{SM}, we extend the standard non-Hermitian topological band theory to nonlocal operators in continuous platforms \cite{Silveirinha_continuous}. 
The detailed analysis shows that the Chern number of the $n$-th band is $\mathcal{C}_{ n}^{+} = s \times (-1)^{n+1} \text{sgn}(\delta) \quad \text{with} \quad \delta = 1 - \left|\rho \right|$, and $s= +/-$ for positive/negative bands \cite{SM}. As previously noted, $\mathcal{C}_n^{-} = - \mathcal{C}_n^{+}$ due to reciprocity. Remarkably, the critical values $\rho = \pm 1$ separate distinct topological phases. 
%In particular, all Chern numbers switch sign when the plates of the guide are interchanged ($\rho \to \rho^{-1}$). This makes sense because the parity inversion $(x,y,z) \to -(x,y,z)$ that exchanges the positions of the plates, also changes the pseudospin as $\pm \to \mp$, and consequently the Chern numbers switch sign $\mathcal{C}^{\pm}_{n} \to - \mathcal{C}^{\pm}_{n}$ \cite{SM}.

%\textcolor{purple}{The invariant $\nu \in \mathbb{Z}_2$ of a frequency band of the $\mathcal{P}\mathcal{T}\mathcal{D}$-symmetric guide relates to the Chern number of ``$+$" sector as $\nu = \mathcal{C}^{+} \, \text{mod} \, 2$ \cite{z2_spin_chern_number}.}

\begin{figure*}[t]
\includegraphics[width=1.\textwidth,clip]{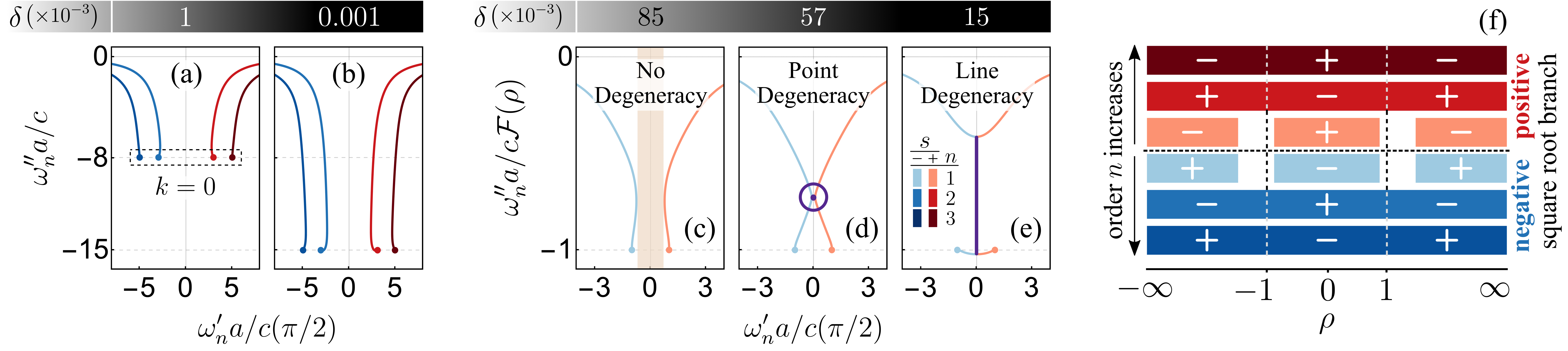}
\caption{\label{fig::phase.transition} (a, b) High-order ($n = 2, 3$) frequency bands projected on the complex plane, with (a) $\delta = 1 - |\rho| = 10^{-3}$ and (b) $\delta = 10^{-6}$. The frequencies associated with $k=0$ are represented by solid dots. (c, d, e) First-order branches ($n = 1$) near the critical resistivity $\rho = 1$  ($\delta \to 0^+$). %(c) A gap separates the positive and negative bands when $\delta \approx 85 \times 10^{-3}$. It is shown as a beige vertical strip. (d) At $\delta \approx 57 \times 10^{-3}$ ($\rho = 0.943$), the bands intersect at an isolated point %$\omega_n = - 2.55\, i\, c/a$ for $k \approx 1.36/a$. The gap closes. 
%(e) The intersection point widens to a line over the imaginary frequency axis ($\delta = 15 \times 10^{-3}$). 
(f) Topological phase diagram. The values of $\mathcal{C}_{n}^{+}$ are given for the range of $\rho$ where they are defined. 
The gap is closed in the white-shaded regions, near $\rho = \pm 1$.
}
\end{figure*}

Figures \hyperref[fig::phase.transition]{2(a)} and \hyperref[fig::phase.transition]{2(b)} show that high-order photonic branches $(n = 2,3,...)$ remain disconnected in the vicinity of $\rho = 1$. Yet, as $\rho \to 1^{-}$, their imaginary parts descend along the imaginary frequency axis. At the critical resistivity $\rho = 1$, the bands diverge, 
%and 
touch at infinity, and their topological charges switch sign. The evolution of the $n=1$ branches is distinct: the gap between the positive and negative frequency spectra is initially open [Fig. \hyperref[fig::phase.transition]{2(c)}], but it closes as $\rho \to 1^{-}$. Specifically, when $\rho = 0.943$, the positive and negative $n = 1$ bands intersect at a single point $\omega_n = - 2.55\, i\, c/a$ for $k = 1.36/a$ [Fig. \hyperref[fig::phase.transition]{2(d)}].
As the resistivity further increases, this degeneracy extends to a line along the imaginary frequency axis [Fig. \hyperref[fig::phase.transition]{2(e)}]. The band gap then  reopens for $\rho>0.943^{-1}$. This specific resistivity range, where the gap remains closed, is dictated solely by the planar geometry of the system \cite{SM}.

%The behaviour of the $n=1$ bands when the critical resistivity is approached from above $\rho \to 1^{+}$ is identical to the one we just described. Indeed, the plots of Fig. \ref{fig::phase.transition} are the same if we switch the sign of $\delta$, because the band structure is unchanged under the transformation $\rho \to \rho^{-1}$ ($\rho = 1 - \delta$ with $\delta \sim 0$). Therefore, the bandgap is closed for $0.943 \leq \rho \leq 0.943^{-1}$ and open otherwise ($\rho > 0$). This range of resistivities is determined uniquely by the planar geometry of the plates and is independent of the waveguide width $a$ \cite{SM}. Evidently, the closing and reopening of the gap mediates the exchange of topological charge between the positive and negative parts of the frequency spectrum, as previously discussed for nonreciprocal Chern insulators \cite{magnetoplasmon, sylvain_link}. The PPW undergoes a topological phase transition after the range of resistivities $0.943 \leq \rho \leq 0.943^{-1}$ is traversed and all Chern  numbers switch sign. We remark that the transitions develop in similar ways around the points $\rho = \pm 1$ because $\omega_n (-\rho) = {\omega}^*_n(\rho)$. 

The gapped phases of the PPW are linked by an intermediate series of exceptional points (EPs) \cite{exceptional_points, exceptional_points_2, eps1, eps2, eps3}. These EPs emerge when the $n = 1$ bands intersect over the imaginary axis ($\omega_n ' = 0$), resulting in the coalescence of the corresponding pseudospinors. Owing to the system cylindrical symmetry, the EPs form a annulus ring. 
As the resistivity nears its critical value, $\delta \to 0$, the decay rate of the EPs diverges logarithmically $\omega_{n}'' \sim (c/a) \ln |\delta|^{-1}$. Thus, the topological phase transition is marked by enhanced absorption, a phenomenon that parallels other photonic systems \cite{exotic_EPS_1, exotic_EPS_2, exotic_EPS_3, exotic_EPS_4, exotic_EPS_5, exotic_EPS_6, exotic_EP_2, exotic_EP_1_top, exotic_EP_3_top, non_hermitian_theory, non_hermitian_poss, top_open_1, top_open_2, top_open_3, top_open_5, fractional_charge}.

%For most frequencies in the complex lower half-plane, the time variation of an excitation is a damped oscillation ($e^{-i \omega_n' t} e^{- \omega_n'' t}$). The decay rate is determined by the imaginary part of the frequency, and is thus a fingerprint of non-Hermitian electrodynamics. Because the $n=1$ frequencies coalesce over the imaginary axis, the exceptional modes do not even oscillate in time ($\omega_n ' = 0$). Besides, their decay rate diverges logarithmically $\omega_{n}'' \sim (c/a) \ln |\delta|^{-1}$ when $k \sim 0$ and the critical resistivity is approached $\delta \to 0$. Heuristically, this boost is due to the impedance match between the dielectric region of the guide and the plates ($\rho \to 1$). We see that the EPs are associated with both the topological phase transition and some form of enhanced electromagnetic response, as in other photonic systems \cite{exotic_EPS_1, exotic_EPS_2, exotic_EPS_3, exotic_EPS_4, exotic_EPS_5, exotic_EPS_6, exotic_EP_2, exotic_EP_1_top, exotic_EP_3_top, non_hermitian_theory, non_hermitian_poss, top_open_1, top_open_2, top_open_3, top_open_5, fractional_charge}. 

Figure \hyperref[fig::phase.transition]{2(f)} presents the detailed topological phase diagram for $\mathcal{C}_{n}^{+}$.
Intriguingly, at a given resistivity $\rho$, the Chern numbers of different bands alternate between
$+1$ and $-1$. 
According to the principle of bulk-edge correspondence, the gap Chern number is correlated with the net number of reflectionless states that propagate at a material interface \cite{review_Ling_Lu, review_Kim, Silveirinha_proof_bulk_edge}. The gap Chern number is determined by the sum of the contributions from all bands below the gap.
For instance, for a gapped system with $-1<\rho<1$, the gap Chern number for the pseudospin ``$+$'' is given by  $\mathcal{C}^{+}_{\text{gap}} = -1+1-1 \dots$.  
This sequence is notably non-convergent, a feature that is quite unique in the realm of topological physics. This property can be attributed to the particle-hole symmetry characteristic of photonic systems \cite{filipa_ill_defined, SM}. Indeed, this symmetry implies that the band diagram comprises an infinite number of bands below the zero-frequency gap, resulting in the non-convergent series. It is important to contrast this with condensed matter systems, where the total topological charge is always finite due to the existence of a well-defined ground state.

%This implies that, while the topology of the individual bands is well-defined, the topology of the band gap itself is not. For instance, for gapped systems with $-1<\rho<1$, the gap Chern number for the pseudospin ``$+$'' is determined by the non-convergent series $\mathcal{C}^{+}_{\text{gap}} = -1+1-1 \dots$. The root of this ill-definition is the particle-hole symmetry of photonic systems \cite{filipa_ill_defined, SM}: because the dispersion diagram is mirrored about the imaginary frequency axis, there are infinitely many bands below the zero-frequency gap. The gap Chern number is the sum of the Chern numbers of those individual bands, so it is given by a non-convergent infinite sum of integers.

%The bulk-edge correspondence relates the Chern invariants of two bulk materials to the net number of reflectionless states propagating at an interface between them \cite{review_Ling_Lu, review_Kim, Silveirinha_proof_bulk_edge}. This way, it conveys physical meaning to nontrivial topological features of photonic band structures. 

\begin{figure*}[t]
\includegraphics[width=1.\textwidth]{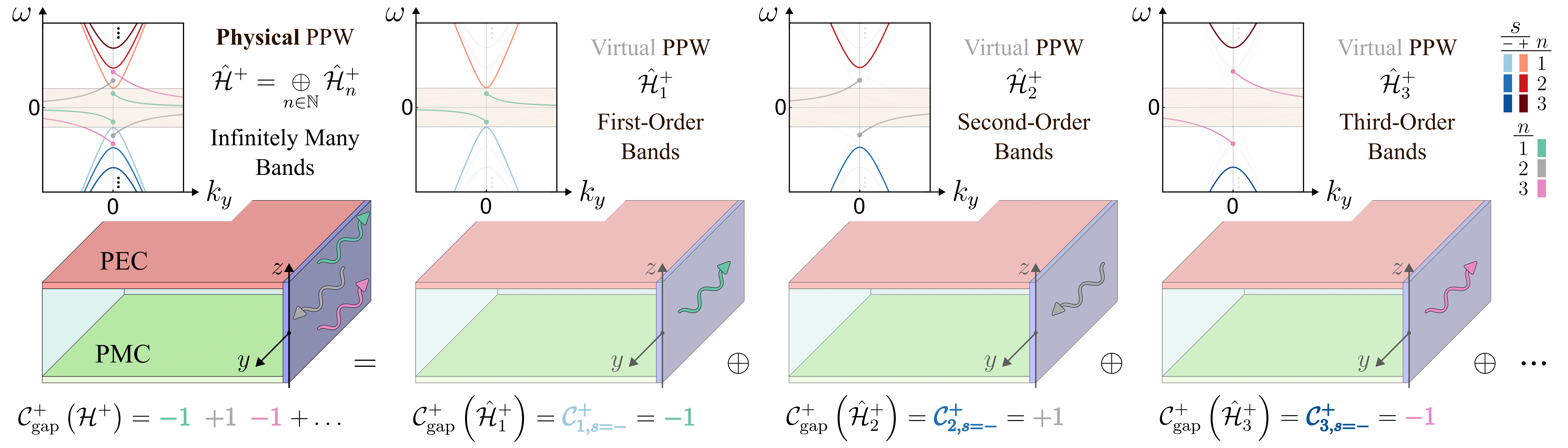}
\caption{\label{fig::waveguide_decomposition} 
Conceptualization of the physical guide terminated with the lateral wall as a juxtaposition of infinite virtual guides. For the pseudospin ``$+$", the Hamiltonian $\hat{\mathcal{H}}^{+}$ is the infinite direct sum of partial Hamiltonians $\hat{\mathcal{H}}^{+}_{n}$. Positive/Negative bulk modes are shown in red/blue tones. The edge waves associated with the quantum numbers $n=1,2,3$ are marked in green, gray, and pink, respectively. The edge modes are shown pictorially with arrows along the lateral wall and are gapless. The details of the simulations and analytical derivations can be found in the SM \cite{SM}.}   

\end{figure*}

To elucidate the implications of the non-convergent series, we next turn our attention to the edge states at the system boundary and their association with the topological charge. For simplicity, our analysis is concentrated on the $\rho=0$ case. This is because non-Hermiticity is known to impact \cite{exceptional_points, non_hermitian_theory, edge_states_non_hermitian_1, edge_states_non_hermitian_2, edge_states_non_hermitian_3}, and in some cases, even challenge \cite{violation_bec_1, violation_bec_2, violation_bec_3, violation_bec_4} the bulk-edge correspondence. Remarkably, under the Hermitian condition, the label ``$n$'' that identifies the bulk bands continues to be a valid quantum number \cite{SM}. This remains true even when the waveguide is closed with a 
$\mathcal{P}\mathcal{T}\mathcal{D}$-symmetric lateral vertical wall with an arbitrary contour. Thus, the physical waveguide can be conceptualized as a juxtaposition of infinitely many uncoupled virtual waveguides, as sketched in Fig. 
\hyperref[fig::waveguide_decomposition]{3}.
All the virtual waveguides share a common band-gap that separates the positive and negative frequency bands. Each bulk virtual guide describes a degenerate two-band Kane-Mele type model \cite{kane_mele_z2_qsh}. 

Focusing our attention in the ``+'' spin polarized waves, the previous discussion shows that the operator $\hat{\mathcal{H}}^{+}$ can be written as a direct sum of the Hamiltonians associated with the virtual guides: $\hat{\mathcal{H}}^{+} = \hat{\mathcal{H}}^{+}_1 \oplus \hat{\mathcal{H}}^{+}_2 \oplus \dots$ [see Fig. \ref{fig::waveguide_decomposition}]. The operator $\hat{\mathcal{H}}^{+}_n$ has a single band below the gap, and has a well-defined topology determined by the ``charge'' $\mathcal{C}_{n}^+=(-1)^n$. Thus, the bulk-edge correspondence implies that each virtual guide supports exactly one ``+'' polarized unidirectional edge state. The direction of propagation of the
%an 
edge state of $\hat{\mathcal{H}}_{n}^{+}$ is strictly locked to the sign of the respective gap Chern number \cite{Silveirinha_proof_bulk_edge}. 
In the supplementary information, we present closed analytical formulas for the edge states supported by a straight lateral wall \cite{SM}. Consistent with the bulk-edge correspondence, each virtual guide accommodates a single gapless edge state (with ``+'' polarization), with the direction of propagation dependent on the parity of $n$ (see Fig. \hyperref[fig::waveguide_decomposition]{3}) \cite{SM}. Therefore, the physical guide hosts an infinite number of gapless scattering-immune edge-states, each corresponding to a term in the ill-defined series $\mathcal{C}^{+}_{\text{gap}} = -1 + 1- 1 + \dots$. In this example, the edge-states are backward waves \cite{SM}.

In summary, we introduced a novel class of non-Hermitian, 
$\mathcal{P}\mathcal{D}$-symmetric, and reciprocal photonic insulators. Our findings demonstrate that, despite the non-Hermitian nature of these systems, Maxwell's equations can be decoupled into spin-up and spin-down states, interconnected by 
$\mathcal{P}\mathcal{D}$-symmetry. Notably, the Chern numbers associated with one of these spin sectors are non-zero, revealing a non-trivial topological characteristic within the sector.
%even though the system as a whole remains topologically trivial due to reciprocity.
We identified two distinct topological phases, differentiated by the degree of material dissipation. The transition between these phases is characterized by the merging of the positive and negative frequency spectra, leading to the formation of a ring of EPs. This phenomenon is particularly significant as it is associated with resonant energy absorption when the impedance of the plates matches that of the dielectric. 
Moreover, our study uncovered a unique topological phase characterized by a non-convergent gap Chern number. We elucidated that this peculiar behavior stems from the particle-hole symmetry. By correlating each term of the divergent series with a topologically protected edge state at the system boundary, we have provided a precise interpretation to this otherwise ill-defined series.

\begin{acknowledgments} 
This work is partially supported by the IET under the A F Harvey Engineering Research Prize, by the Simons Foundation under the award 733700 (Simons Collaboration in Mathematics and Physics, “Harnessing Universal Symmetry Concepts for Extreme Wave Phenomena”), by Fundação para a Ciência e a Tecnologia and Instituto de Telecomunicações under project UIDB/50008/2020 and by FCT-Portugal through Grant No. CEECIND/07471/2022.

\end{acknowledgments} 
%TC:endignore

% The \nocite command causes all entries in a bibliography to be printed out
% whether or not they are actually referenced in the text. This is appropriate
% for the sample file to show the different styles of references, but authors
% most likely will not want to use it.

%\nocite{*}

\bibliography{article}% Produces the bibliography via BibTeX.

\end{document}

%I don't know what is this
\preprint{APS/123-QED}

\title{Supplementary Material for ``Non-Hermitian Photonic Spin-Hall Insulators''}% Force line breaks with \\

\author{Rodrigo P. Câmara}
\affiliation{Instituto de Telecomunicações, Instituto Superior Técnico, University of Lisbon, Avenida Rovisco Pais 1, Lisboa, 1049001 Portugal}

\author{Tatiana G. Rappoport}
\affiliation{Instituto de Telecomunicações, Instituto Superior Técnico, University of Lisbon, Avenida Rovisco Pais 1, Lisboa, 1049001 Portugal}	
\affiliation{Instituto de F\'\i sica, Universidade Federal do Rio de Janeiro, C.P. 68528, 21941-972 Rio de Janeiro RJ, Brazil}

\author{M\'{a}rio G. Silveirinha}
\affiliation{Instituto de Telecomunicações, Instituto Superior Técnico, University of Lisbon, Avenida Rovisco Pais 1, Lisboa, 1049001 Portugal}

\date{\today}% It is always \today, today,
             %  but any date may be explicitly specified

%\keywords{Suggested keywords}%Use showkeys class option if keyword

\maketitle

\section{\label{sec::pseudospin_polarization_in_PD_symmetric_media} Pseudospin-Polarization in $\mathcal{P}\mathcal{D}$-Symmetric Reciprocal Media}

We consider reciprocal electromagnetic systems formed by isotropic materials characterized by the relative permittivity $\epsilon$ and permeability $\mu$. For these platforms, the frequency-domain (source-free) Maxwell's equations read

\begin{equation}
    \bm{M}^{-1} \cdot
    \begin{pmatrix} 
    \bm{0}_{3\times3} & i c \nabla \times \bm{1}_{3\times 3} \\ 
    - i c \nabla \times \bm{1}_{3\times 3} & \bm{0}_{3\times3} 
    \end{pmatrix} \cdot 
    \bm{f} = \omega
    \bm{f}
    \quad  \text{with} \quad 
    \bm{M}(\bm{r}) = 
    \begin{pmatrix}
    \epsilon(\bm{r}) \bm{1}_{3\times 3} & \bm{0}_{3\times3} \\
    \bm{0}_{3\times3} & \mu(\bm{r}) \bm{1}_{3\times 3}
    \end{pmatrix} ,
    \label{eq::maxwell_equations}
\end{equation}

\noindent where $c$ is the speed of light in vacuum, $\omega$ is the oscillation frequency, $\bm{r}= (x, y, z)^\intercal$ is the position vector in a Cartesian reference frame and $\bm{f} = \left( \bm{E}, Z_0 \bm{H} \right)^\intercal$ is a $6$-vector built from the components of the electromagnetic fields normalized to the same units by means of the vacuum impedance $Z_0$. The symbol $\cdot$ denotes the matrix multiplication operation, $^\intercal$ is the transposition operator and $\bm{1}_{3\times3}$/$\bm{0}_{3\times3}$ are the $3\times3$ identity/null matrices. 

Let us consider that the permittivity and permeability are related as

\begin{equation}
    \epsilon(x,y,z) = \mu(x,y,-z).
    \label{eq::symmetry}
\end{equation}

\noindent Then, the eigenvalue problem in Eq.(\ref{eq::maxwell_equations}) can be divided into two independent sets of equations \cite{Chen2015}

\begin{equation}
    \pm \frac{i c}{\epsilon (z)}
    \begin{pmatrix}
    0 & \partial_z & \partial_y \\
    - \partial_z & 0 & - \partial_x \\
    \partial_y & - \partial_x & 0
    \end{pmatrix} \cdot
    \bm{\Psi}^\pm(-z) = \omega \bm{\Psi}^\pm(z) \quad \text{with} \quad \bm{\Psi}^\pm (z) =
    \begin{pmatrix}
    E_x(z) \mp Z_0 H_x(-z) \\
    E_y(z) \mp Z_0 H_y(-z) \\
    E_z(z) \pm Z_0 H_z(-z)
    \end{pmatrix},
    \label{eq::decoupled_maxwell_equations}
\end{equation}

\noindent formally equivalent to two decoupled eigenvalue problems $\hat{\mathcal{H}}^\pm \bm{\Psi}^\pm (z) = \omega \bm{\Psi}^\pm(z)$. A generic point of space $(x,y,z)$ is denoted as $(z)$ and its mirror-symmetric counterpart $(x,y,-z)$ as $(-z)$ for conciseness. Equation (\ref{eq::decoupled_maxwell_equations}) is nonlocal in space because the left and right hand sides are evaluated at mirror symmetric points. The pseudospinors $\bm{\Psi}^+$ and $\bm{\Psi}^-$ form a basis of solutions of Eq. (\ref{eq::maxwell_equations}) which follows from the symmetry of the guide: the electric and parity-transformed magnetic fields are in/out-of phase for the ``$+$''/``$-$'' class. It is important to note that the "pseudospin" degree of freedom is unrelated to the angular momentum (orbital or spin) of the electromagnetic field. It rather refers to the non-local polarization determined by the $\mathcal{P}\mathcal{D}$-symmetry operator. This polarization is the internal degree of freedom that can still convey a topologically nontrivial character to the guide, even though reciprocity renders the global Chern topology trivial. Importantly, the pseudospin decomposition remains valid even when the permittivity and permeability are frequency dependent and complex-valued.

Let $\hat{\mathcal{P}}$ be the parity (mirror) transformation that flips the $z$-spatial coordinate and $\hat{\mathcal{D}}:\left(\bm{E},\bm{H}\right) \to \left(Z_0 \bm{H},-\bm{E}/Z_0\right)$ be a duality mapping that exchanges the role of the two fields. Both $\hat{\mathcal{P}}$ and $\hat{\mathcal{D}}$ are operators that map the $\bm{f}$-space into itself as $\hat{\mathcal{P}}: \bm{f}(z) \to \mathcal{P} \cdot \bm{f}(-z)$ and $\hat{\mathcal{D}}: \bm{f}(z) \to \mathcal{D} \cdot \bm{f}(z)$ where \cite{Silveirinha_ptd}

\begin{equation}
    \mathcal{P} =
    \begin{pmatrix}
    \bm{V} & \bm{0}_{3\times3} \\
    \bm{0}_{3\times3} & -\bm{V}
    \end{pmatrix} \quad \text{with} \quad
    \bm{V} = \text{diag}\left(1,1,-1\right) \quad \text{and} \quad 
    \mathcal{D} =
    \begin{pmatrix}
    \bm{0}_{3\times3} & \bm{1}_{3 \times 3} \\
    - \bm{1}_{3 \times 3} & \bm{0}_{3\times3}
    \end{pmatrix}.
    \label{eq::transformations_matrices}
\end{equation}

\noindent The composition operator

\begin{equation}
    \hat{\mathcal{P}} \cdot \hat{\mathcal{D}}:
    \begin{pmatrix}
        \bm{E}(z) \\
        Z_0 \bm{H}(z)
    \end{pmatrix} \to
    \begin{pmatrix}
        Z_0 \bm{V} \cdot \bm{H}(-z) \\
        \bm{V} \cdot \bm{E}(-z)
    \end{pmatrix}
    \label{eq::pd_transformation}
\end{equation}

\noindent maps the Cartesian components of the electric and magnetic fields as $E_z (z) \to -Z_0 H_z (-z)$, $H_z (z) \to -E_z (-z)/Z_0$, $E_{x,y} (z) \to + Z_0 H_{x,y} (-z)$ and $H_{x,y} (z) \to + E_{x,y} (-z) /Z_0$. It is straightforward to show that for reciprocal isotropic dielectrics the Maxwell's equations (\ref{eq::maxwell_equations}) are $\hat{\mathcal{P}} \cdot \hat{\mathcal{D}}$ symmetric if and only if the permittivity and permeability are linked as in Eq. (\ref{eq::symmetry}).

For $\hat{\mathcal{P}} \cdot \hat{\mathcal{D}}$ symmetric reciprocal systems, if $\bm{f}$ is a solution of the Maxwell's equations then $\hat{\mathcal{P}} \cdot \hat{\mathcal{D}} \cdot \bm{f}$ also is. Because the linear operator $ \hat{\mathcal{P}} \cdot \hat{\mathcal{D}}$ satisfies $\left[ \hat{\mathcal{P}} \cdot \hat{\mathcal{D}} \right] ^2 = {{{\bf{1}}_{6 \times 6}}}$, the eigenfunctions of the system (\ref{eq::maxwell_equations}) can be split into two subsets such that $\hat{\mathcal{P}} \cdot \hat{\mathcal{D}} \cdot \bm{f} = \pm \bm{f}$. Clearly, $\bm{\Psi}^{\pm}$ coincides with the first 3 components of the six-vector $\bm{f} \mp \hat{\mathcal{P}} \cdot \hat{\mathcal{D}} \cdot \bm{f}$. This property implies that the pseudospinors $\bm{\Psi}^{\pm}$ describe the dynamics of the electric field associated with waves with the symmetry $\hat{\mathcal{P}} \cdot \hat{\mathcal{D}} \cdot \bm{f} = \mp \bm{f}$. Thereby, each pseudospinor symmetry determines itself a solution of the Maxwell's equations with the electric and magnetic fields given by

\begin{equation}
    \begin{pmatrix}
        \bm{E} (z) \\
        Z_0 \bm{H} (z)
    \end{pmatrix} = 
    \begin{pmatrix}
        \bm{\Psi}^{\pm} (z) \\
        \mp \bm{V} \cdot \bm{\Psi}^{\pm} (-z)
    \end{pmatrix}.
\label{eq::relation_fields_pseudospinors}
\end{equation}

In the lossless case, when both the permittivity and permeability are real-valued, Eq. (\ref{eq::maxwell_equations}) is time-reversal ($\hat{\mathcal{T}}$) invariant. In such a case, the system becomes $\hat{\mathcal{P}} \cdot \hat{\mathcal{T}} \cdot \hat{\mathcal{D}}$ invariant \cite{Silveirinha_ptd}. The $\hat{\mathcal{P}} \cdot \hat{\mathcal{T}} \cdot \hat{\mathcal{D}}$ operator is anti-linear and satisfies 
$\left[ \hat{\mathcal{P}} \cdot \hat{\mathcal{T}} \cdot \hat{\mathcal{D}} \right] ^2 = -{{{\bf{1}}_{6 \times 6}}}$. Such a formal property allows one to establish a perfect parallelism with the spin-Hall effect and regard the system as a $\mathbb{Z}_2$ photonic insulator \cite{Silveirinha_ptd}. The $\mathbb{Z}_2$ index is written in terms of the topological charge ${\mathcal{C}}^+  =  - \mathcal{C}^{-}$ of the spin operators $\hat {\mathcal{H}}^ \pm$ introduced in the main text. In contrast, in the non-Hermitian case the time-reversal symmetry is broken. Yet, as the system remains reciprocal, it is still possible to guarantee that the two spin operators have opposite topological charge (${\mathcal{C}}^+  =  - \mathcal{C}^{-}$), exactly as in the Hermitian case. In this sense, reciprocal $\hat{\mathcal{P}} \cdot \hat{\mathcal{D}}$ systems may be regarded as generalized (non-Hermitian) $\mathbb{Z}_2$-photonic insulators.

\section{\label{sec::boundary_conditions_in_the_ppw} Boundary Conditions for the PPW}

The permittivity and permeability of air are related as in Eq. (\ref{eq::symmetry}) ($\epsilon(z) = \mu(-z) = 1$). In the parallel-plate waveguide (PPW) schematized in Fig. 1 of the main text, the pseudospin-decomposition can be preserved by choosing plates that are $\mathcal{P}\mathcal{D}$-symmetric. This ensures that the full guide is invariant under a $\mathcal{P}\mathcal{D}$ transformation and that the decoupling in Eq. (\ref{eq::decoupled_maxwell_equations}) remains valid [see discussion in Sec. \ref{sec::pseudospin_polarization_in_PD_symmetric_media}]. The plates are characterized by a (Leontovich) boundary condition of the form

\begin{equation}
\begin{cases}
    - Z_+ \, \uvec z \times \bm{H}_{\text{tan}}(a/2) = \bm{E}_{\text{tan}} (a/2)\\
    Z_- \, \uvec z \times \bm{H}_{\text{tan}}(-a/2) = \bm{E}_{\text{tan}} (-a/2)
\end{cases},
\label{eq::boundary_conditions}
\end{equation}

\noindent where $Z_{i}$ is the surface impedance of the $i=+/-$ (top/bottom, respectively) plate, $\bm{E}_{\text{tan}} = \bm{E} - \uvec{z} \left( \uvec{z} \cdot \bm{E} \right)$ is the tangential electric field, $a$ is the distance between the plates, $\times$ denotes the cross product between vectors and $\uvec{z}$ is the unitary vector along the positive $z$-spatial direction. Under a $\mathcal{P}\mathcal{D}$-transformation, the boundary conditions (\ref{eq::boundary_conditions}) become 

\begin{equation}
\begin{cases}  
    -\left(Z_0^2/ Z_- \right) \uvec z \times \bm{H}_{\text{tan}} (a/2) = \bm{E}_{\text{tan}}(a/2) \\
    \left(Z_0^2/Z_+\right) \uvec z \times \bm{H}_{\text{tan}} (-a/2) = \bm{E}_{\text{tan}}(-a/2)
\end{cases}
\label{eq::boundary_conditions2}
\end{equation}

\noindent We used the transformation rules for the fields: $\left(\bm{E}_{\text{tan}} (z), \bm{H}_{\text{tan}} (z)\right) \to \left(+Z_0 \bm{H}_{\text{tan}} (-z), +\bm{E}_{\text{tan}} (-z)/Z_0 \right)$, in agreement with Eq. (\ref{eq::pd_transformation}). $\mathcal{P}\mathcal{D}$-invariance requires that (\ref{eq::boundary_conditions}) and (\ref{eq::boundary_conditions2}) must coincide. This is possible when 

\begin{equation}
    Z_{+} Z_{-} = Z_{0}^{2}.
    \label{eq::symmetry_boundary_conditions}
\end{equation}

\section{\label{sec::bulk_modes} Bulk Modes}

To find the pseudospinor-polarized states $\bm{\Psi}^\pm$ that propagate in the $\mathcal{P}\mathcal{D}$-symmetric PPW [see Eq. (\ref{eq::decoupled_maxwell_equations})], we first determine the electromagnetic fields $\bm{E}$ and $\bm{H}$ that solve Eq. (\ref{eq::maxwell_equations}) subject to the boundary conditions in Eq. (\ref{eq::boundary_conditions}). 

\begin{comment}
Following the discussions in Secs. \ref{sec::pseudospin_polarization_in_PD_symmetric_media} and \ref{sec::boundary_conditions_in_the_ppw}, we aim to divide the optical states that propagate inside the PD-symmetric guide into the two pseudospin classes $\bm{\Psi}^\pm$. This goal can be achieved by first determining the electromagnetic fields $\bm{E}$ and $\bm{H}$ that solve Eq.(\ref{eq::maxwell_equations}) subject to the boundary constraints in Eq.(\ref{eq::boundary_conditions_PD_symmetric}) and then combining them according to the recipe in Eq.(\ref{eq::decoupled_maxwell_equations}) that yields the up and down spinors. Here, we consider transverse electric (TE) modes as a starting point.
\end{comment}

\begin{comment}
The physically relevant bulk modes might be decomposed into transverse electric (TE) and transverse magnetic (TM) field configurations \textcolor{blue}{(justify why no TEM?)} but only one group is necessary to generate the complete set of corresponding solutions in the pseudospin basis. In fact, a TE mode is mapped into a TM one under a PD-transformation and we saw that such change preserves the spinor states apart from an overall sign [see Eq.(\ref{eq::action_parity_duality}) and succeeding discussion in Sec. \ref{sec::pseudospin_polarization_in_PD_symmetric_media}].
\end{comment}

 Due to the invariance of the guide under continuous translations in the $xoy$ plane, the fields can be factorized as $\bm{E}(\bm{r}, t) = \bm{E}_{\bm{k}} (z) e^{-i \omega t} e^{i \bm{k} \cdot \bm{r}}$ and $\bm{H}(\bm{r}, t) = \bm{H}_{\bm{k}} (z) e^{-i \omega t} e^{i \bm{k} \cdot \bm{r}}$, where $\bm{k}$ is the in-plane real-valued wave vector ($\bm{k} \cdot \uvec{z} = 0$) with magnitude $k$. For convenience, first we choose the orientation of the Cartesian reference frame such that $\hat{\bm{x}} \cdot \hat{\bm{k}} = 1$. 
% and $\uvec{z}$ is directed along the axis shown in Fig. 1 of the main text. 

As a starting point, we consider the transverse electric (TE) modes. The \textit{ansatz} $\bm{H}_{\bm{k}}(z) = h(z) \bm{\hat{x}} + \alpha \partial_z h (z) \bm{\hat{z}}$ describes a TE field configuration, where $\alpha$ is some coefficient to be determined. The divergence law for magnetism $\nabla \cdot \bm{H}(\bm{r}, t) = 0$ yields the relation $ikh(z) + \alpha \partial_z^2 h(z) = 0$. On the other hand, it follows from the Helmholtz equation $\nabla^2 H_x + \omega^2 H_x/c^2 = 0$ that $\partial^2_z h(z) + (\omega^2/c^2 - k^2) h(z) = 0$. Combining both equations one finds that $\alpha \equiv ik/\kappa^2$, where we define the transverse wavenumber $\kappa \equiv \sqrt{\omega^2/c^2 - k^2}$. Using Ampère's law $\nabla \times \bm{H} = -i \omega \epsilon_0 \bm{E}$, one easily finds that the electric field is given by $\bm{E}_{\bm{k}}(z) = i \frac{\mu_0 \omega}{\kappa^2} \partial_z h(z) \hat{\bm{y}}$. 

The differential equation $\partial_z^2 h(z) + \kappa^2 h(z) = 0$ has the general solution $h(z) = A e^{i \kappa z} + B e^{- i \kappa z}$, with $A$ and $B$ arbitrary constants. By imposing the boundary conditions (\ref{eq::boundary_conditions}), we find that the two coefficients are constrained by:

\begin{equation}
    \bm{Q} \cdot
    \begin{pmatrix}
    A \\
    B
    \end{pmatrix} = 0 \quad \text{with} \quad
    \bm{Q} = 
    \begin{pmatrix}
    - \left(\omega/c \kappa - \rho \right) e^{i \kappa a/2} & \left( \omega/c \kappa + \rho \right)e^{- i \kappa a/2} \\
    \left(1 + \rho \omega/c \kappa \right) e^{- i \kappa a/2} & \left(1 - \rho \omega/c \kappa \right)e^{i \kappa a/2}
    \end{pmatrix}.
    \label{eq::boundary_conditions_AB_result}
\end{equation}

\noindent We used $Z_{\pm} = \rho^{\pm 1} Z_0$ with $\rho$ real-valued in conformity with Eq. (\ref{eq::symmetry_boundary_conditions}). Non-trivial solutions of the homogeneous linear system must satisfy the modal condition $\text{det}\,(\bm{Q}) = 0$, i.e.,

\begin{equation}
    e^{2 i \kappa a} = \frac{\omega + c \rho \kappa}{\omega - c \rho \kappa} \frac{\rho \omega + c \kappa}{\rho \omega - c \kappa},
    \label{eq::modal_equation}
\end{equation}

\noindent where $\omega = s \times c \sqrt{k^2 + \kappa^2}$ with the square root branch determined by $s=\pm$. This result shows that the transverse wavenumber is generally a function of $k$, $\rho$ and $a$. 
%Since distinct values of the guide width $a$ and $a'$ lead to differently scaled versions of the the transverse wavenumber as $\kappa(k; \rho, a') = (a/a')\, \kappa((a/a')k; \rho, a)$, we analyse the system for a fixed $a$ henceforward. 

In the limit $k \to 0^+$, Eq. (\ref{eq::modal_equation}) is analytically solvable and has infinitely many solutions $\kappa_n^{(s = \pm)}$ indexed by $n = 1,2,\dots$ for both branches $s = \pm$ of the square root [see Sec. \ref{sec:important_limits}]. For $0<k<\infty$, however, Eq. (\ref{eq::modal_equation}) needs to be numerically solved. To this end, we use the Nelder-Mead minimization scheme \cite{Nelder1965}. The Nelder-Mead algorithm is suited to find local minima of a scalar field $f: \mathbb{C} \to \mathbb{R}$ \cite{Nelder1965}. It starts with a set of three points $p_0$, $q_0$, and $s_0$ in the complex plane, that form a triangle $\triangle p_{0}q_{0}s_{0}$ with centroid $c_0$. Then, a composition of $N$ transformations – reflections, expansions, etc. – is applied to the initial triangle, depending on how the values of the function $f$ at the vertices evolve. This produces a series of triangles $\left\{ \triangle p_{n} q_{n} s_{n} \right\}_{n}$ with centroids $\left\{ c_{n} \right\}_{n}$ ($n=0,\dots,N$), such that $c_{n}$ converges to a local minimum. For this reason, the Nelder-Mead method is also called the downhill simplex method. In our case, we take the function 

\begin{equation}
    f_k : \mathbb{C} \to \mathbb{R}^{+}_{0},\, \kappa \mapsto f_k (\kappa) = \left| e^{2 i \kappa a} - \frac{\omega (k, \kappa) + c \rho \kappa}{\omega (k, \kappa) - c \rho \kappa} \frac{\rho \omega (k, \kappa) + c \kappa}{\rho \omega (k, \kappa) - c \kappa} \right|,
\label{eq::function_to_minimize}
\end{equation}

\noindent with $\kappa$ the unknown parameter. In the above, $\omega = \omega (k, \kappa)$ represents the function $\omega = s \times c \sqrt{k^2 + \kappa^2}$.
%which is parametrically defined in terms of the in-plane momentum magnitude $k$ and has the complex-valued transverse wavenumber $\kappa$ as argument. Here, we write $\omega = \omega (k, \kappa)$ to emphasize that the frequency values are written in terms of the momenta: $\omega = s \times c \sqrt{k^2 + \kappa^2}$. By construction, the zeros of the function $f_k$ are the solutions $\kappa_n (k)$ of the modal constraint in Eq. (\ref{eq::modal_equation}) evaluated at $k$. Besides, $f_k$ is non-negative, so its zeros are also minima. This means that the application of the Nelder-Mead minimization algorithm to the function $f_k$ in Eq. (\ref{eq::function_to_minimize}) yields one transverse wavenumber solution $\kappa_n (k)$, provided that the initial triangle $\triangle p_{0}q_{0}s_{0}$ is small enough and close enough to that value. Once we now $\kappa_n (k)$, we can generate a new triangle $\triangle p_{0}q_{0}s_{0}$ in the complex plane with centroid $\kappa_n (k)$ and feed it to the Nelder-Mead scheme applied to the function $f_{k + \Delta k}$, for small $\Delta k$. The algorithm converges to the solution $\kappa_n (k + \Delta k)$ of the modal equation (\ref{eq::modal_equation}) evaluated at a slightly larger in-plane momentum magnitude. 
We use the Nelder-Meader method recursively, so that  $\kappa$ is calculated for a generic real-valued $k$, starting from the analytical solution at $k = 0$. In other words, we  numerically continue the 
%analytical 
solution at $k=0$ to $k>0$.
%using the Nelder-Meader algorithm.  

From the first identity of the system in Eq. (\ref{eq::boundary_conditions_AB_result}), we find that $B = h_n A$ with $h_n \equiv e^{ i \kappa_n a} \frac{\omega_n - c \rho \kappa_n}{\omega_n + c \rho \kappa_n}$. The components of the electromagnetic fields associated with TE modes are thus

\begin{equation}
    \begin{cases}
    H_{\bm{k}, x} (z) = A \left( e^{i \kappa_n z} + h_n e^{-i \kappa_n z}  \right) \\
    H_{\bm{k}, z} (z) = -A \frac{k}{\kappa_n} \left( e^{i \kappa_n z} - h_n e^{-i \kappa_n z} \right) \\
    E_{\bm{k}, y} (z) = - A \frac{\mu_0 \omega_n}{\kappa_n} \left( e^{i \kappa_n z} - h_n e^{-i \kappa_n z} \right)
    \end{cases}
    \label{eq::field_structure_TE}
\end{equation}

\noindent with $A$ being an arbitrary complex number. Using now the projection (\ref{eq::decoupled_maxwell_equations}), one finds that the pseudospinors associated with the TE modes [Eq. (\ref{eq::field_structure_TE})] can be written as $\bm{\Psi}^{\pm}_{n}(\bm{r}, t) = e^{-i \omega_n t} e^{i \bm{k} \cdot \bm{r}} \bm{\Psi}^{\pm}_{\bm{k}, n}(z)$ with 

\begin{equation}
    \bm{\Psi}^{\pm}_{\bm{k}, n} (z) \propto
    \begin{pmatrix}
    \mp \left( e^{-i \kappa_n z} + h_n e^{i \kappa_n z} \right) \\
    - \frac{\omega_n}{c \kappa_n} \left( e^{i \kappa_n z} - h_n e^{-i \kappa_n z}  \right) \\
    \mp \frac{k}{\kappa_n} \left( e^{-i \kappa_n z} - h_n e^{i \kappa_n z} \right) 
    \end{pmatrix}.
    \label{eq::explicit_pseudospinors}
\end{equation}

\noindent Recall that we picked $\uvec{x} \cdot \uvec{k} = 1$. In the general case, the first component of the pseudospinor corresponds to the projection on $\uvec{k}$, the second component corresponds to the projection on $\uvec{z} \times \uvec{k}$, whereas the third component gives the projection on $\uvec{z}$. For completeness, we remark that the transverse magnetic (TM) modes of the PPW lead to the same pseudospinor decomposition. 

\section{\label{sec::chern_numbers} Spin Chern Numbers}

The electrodynamics in the waveguide is generally not conservative, as there is energy dissipated when $\rho > 0$ and energy pumped into the guide when $\rho < 0$. Non-Hermitian Hamiltonians $\hat{\mathcal{H}} \neq \hat{\mathcal{H}}^\dagger$ describing systems with loss or gain have left $\phi^\text{L}_n$ and right $\phi^\text{R}_n$ eigenstates that, though linked to the same complex eigenenergy $E_n$, are usually independent \cite{non_hermitian_theory}. Here $n$ labels the energy band, $\hat{\mathcal{H}}^\dagger \phi^\text{L}_n = E_n^* \phi^\text{L}_n$ and $\hat{\mathcal{H}} \phi^\text{R}_n = E_n \phi^\text{R}_n$. The extension of topological band theory to complex spectra leads to the definition 
%of four possibly different Berry potentials $\bm{\mathcal{A}}^{\alpha, \beta}_{\bm{k},n} \equiv i \braket{\phi^\alpha_n | \nabla_{\bm{k}} \phi^\beta_n}$ for separable bands, based on distinct combinations of the left and right functions ($\alpha, \beta \in \left\{ \text{L}, \text{R} \right\}$). 
of the Berry potential $\bm{\mathcal{A}}_{\bm{k},n} \equiv i \braket{\phi^L_n | \nabla_{\bm{k}} \phi^R_n}$ for separable bands, based on the combination of the left and right eigenfunctions \cite{non_hermitian_theory}. Here, $\nabla_{\bm{k}}$ represents the surface gradient in momentum space and the Dirac notation indicates a suitable inner product. It is supposed that the eigenstates are normalized as $\braket{\phi^L_m | \phi^R_n} = \delta_{m,n}$. %Strikingly, the Chern numbers associated to each potential are equal \cite{non_hermitian_theory}.

In the PPW, switching loss into gain is equivalent to switch the sign of $\rho$. As a consequence, the operator $\hat{\mathcal{H}}^\dagger$ models a guide with resistivity $-\rho$. This means that $\bm{\Psi}^{\rm{L}}_{n}(z;\rho)=\bm{\Psi}^{\rm{R}}_{n}(z;-\rho)$. Thus, the Berry connection can be written as $\bm{\mathcal{A}}_{\bm{k},n}^{\pm} \equiv i \braket{\bm{\Psi}^\pm_{\bm{k},n}(-\rho) | \nabla_{\bm{k}} \bm{\Psi}^\pm_{\bm{k},n} (\rho)}$. It is built from properly normalized versions $\bm{\Psi}^{\pm}_{\bm{k},n}$ of the pseudospinors in Eq. (\ref{eq::explicit_pseudospinors}) that satisfy $\braket{\bm{{\Psi}}^\pm_{\bm{k}, n} (-\rho)|\bm{{\Psi}}^\pm_{\bm{k}, n} (\rho)} = 1$. We choose $\braket{\bm{F}|\bm{Q}} = \int_{-a/2}^{a/2}dz\, \bm{F}^\dagger (z)\cdot \bm{Q} (z)$ as the inner product of any two vectors $\bm{F}$ and $\bm{Q}$. Note that the Berry potential is independent of the $z$-spatial coordinate.

The Chern number $\mathcal{C}_{n}^{\pm}$ is the topological invariant determined by the integral of the Berry curvature $\nabla_{\bm{k}} \times \bm{\mathcal{A}}^{\pm}_{\bm{k}, n}$ over the momentum space, i.e., the Euclidean plane ($\bm{k}$ is a generic vector in the $xoy$ plane). Stokes theorem tells us that \cite{Silveirinha_continuous}

\begin{equation}
    \mathcal{C}_{n}^{\pm} = (2\pi)^{-1}\left[\lim_{k \to \infty} - \lim_{k \to 0^+}\right] k \int_{0}^{2\pi} d\phi\, \mathcal{A}_{\phi, n}^{\pm} \quad \text{with} \quad \mathcal{A}_{\phi, n}^{\pm} = \bm{\mathcal{A}}_{\bm{k}, n}^{\pm} \cdot \bm{\hat{\phi}},
    \label{eq::chern_numbers_main_text}
\end{equation}

\noindent provided that the Berry potential is smooth for $k \in \left]0,\infty \right[$. Here, $\bm{k} = (k, \phi) \in \left[0, \infty \right[ \times \left[0, 2\pi \right[$ is associated with a system of polar coordinates of the momentum space so that $\nabla_{\bm{k}} = \bm{\hat{k}} \partial_k + \bm{\hat{\phi}} k^{-1} \partial_\phi$. It is implicit that we adopt a gauge such that the eigenfunctions $\bm{\Psi}_{\bm{k}, n}^{\pm}$ are invariant under arbitrary rotations around the $z$-axis, consistent with the continuous rotation symmetry of the PPW. Thus, the components of the states $\bm{\Psi}_{\bm{k}, n}^{\pm}$ in Eq.(\ref{eq::explicit_pseudospinors}) are independent of the $\phi$ coordinate. As a result, $\mathcal{A}_{\phi, n}^{\pm}$ in Eq. (\ref{eq::chern_numbers_main_text}) is also independent of $\phi$, and the Chern number becomes $\mathcal{C}_{n}^{\pm} = \left[\lim_{k \to \infty} - \lim_{k \to 0^+}\right] k \mathcal{A}_{\phi, n}^{\pm}$. Taking into account the normalization of the eigenfunctions, the azimuthal component of the Berry potential $\mathcal{A}_{\phi, n}^{\pm}$ can be written as
%This implies that the overlap $\braket{\bm{\Psi}_{\bm{k},n}^{\pm}(-\rho) | \bm{\Psi}_{\bm{k},n}^{\pm}(\rho)}$ commutes with the partial derivative $\partial_\phi$ and hence the pseudospinors can be correctly normalized for the angular projection of the Berry potential $\mathcal{A}_{\phi, n}^{\pm}$ as

\begin{equation}
    \mathcal{A}_{\phi, n}^{\pm} = i k^{-1} \frac{\braket{\bm{\Psi}_{\bm{k}, n}^{\pm}(-\rho) | \partial_\phi \bm{\Psi}_{\bm{k}, n}^{\pm}(\rho)}}{\braket{\bm{\Psi}_{\bm{k}, n}^{\pm}(-\rho) | \bm{\Psi}_{\bm{k},n}^{\pm}(\rho)}}
    \label{eq::berry_potential_factozied}.
\end{equation}

\noindent Making use of the symmetries $\kappa_n(k; -\rho) = \kappa^*_n(k; \rho)$, $\omega_n(k; -\rho) = \omega^*_n (k; \rho)$ of the modal Eq.(\ref{eq::modal_equation}), the denominator can be determined from Eq. (\ref{eq::explicit_pseudospinors}) as

\begin{equation}
\begin{split}
    \braket{\bm{\Psi}_{\bm{k}, n}^{\pm} (-\rho)|\bm{\Psi}_{\bm{k},n}^{\pm} (\rho)} A^{-2} Z_0^{-2} &= \int_{-a/2}^{a/2}dz\,
    \begin{pmatrix}
    \mp \left( e^{-i \kappa^*_n z} + h^*_n e^{i \kappa^*_n z} \right) \\
    - \frac{\omega^*_n}{c \kappa^*_n} \left( e^{i \kappa^*_n z} - h^*_n e^{-i \kappa^*_n z} \right) \\
    \mp \frac{k}{\kappa^*_n} \left( e^{-i \kappa^*_n z} - h^*_n e^{i \kappa^*_n z} \right)
    \end{pmatrix}^\dagger \cdot
    \begin{pmatrix}
    \mp \left( e^{-i \kappa_n z} + h_n e^{i \kappa_n z} \right) \\
    - \frac{\omega_n}{c \kappa_n} \left( e^{i \kappa_n z} - h_n e^{-i \kappa_n z} \right) \\
    \mp \frac{k}{\kappa_n} \left( e^{-i \kappa_n z} - h_n e^{i \kappa_n z} \right)
    \end{pmatrix}
    \\  
    & = \frac{4 a \omega_n^2}{c^2 \kappa_n^2} + \frac{i k^2}{\kappa_n^3}\left(e^{i\kappa_n a} - e^{-i \kappa_n a}\right) \left( h_n + h_n^{-1}\right).
\end{split}    
\label{eq::denominator_berry_potential}
\end{equation}

\noindent Next, we turn our attention to the numerator of Eq.(\ref{eq::berry_potential_factozied}). We note that $\partial_\phi$ can only act on the vector basis $(\bm{\hat{k}}, \bm{\hat{\phi}}, \bm{\hat{z}})$.
%$(\bm{\hat{x}}, \bm{\hat{y}}, \bm{\hat{z}}) = (\bm{\hat{k}}, \bm{\hat{\phi}}, \bm{\hat{z}})$. 
Using $\partial_\phi \bm{\hat{k}} = \bm{\hat{\phi}}$, $\partial_\phi \bm{\hat{\phi}} = - \bm{\hat{k}}$ and $\partial_\phi \bm{\hat{z}} = 0$, it is found that: 

\begin{equation}
    \begin{split}
        \braket{\bm{\Psi}_{\bm{k}, n}^{\pm} (-\rho)| \partial_\phi \bm{\Psi}_{\bm{k},n}^{\pm} (\rho)} A^{-2} Z_0^{-2} &= \int_{-a/2}^{a/2}dz\,
        \begin{pmatrix}
        \mp \left( e^{-i \kappa^*_n z} + h^*_n e^{i \kappa^*_n z} \right) \\
        - \frac{\omega^*_n}{c \kappa^*_n} \left( e^{i \kappa^*_n z} - h^*_n e^{-i \kappa^*_n z} \right) \\
        \mp \frac{k}{\kappa^*_n} \left( e^{-i \kappa^*_n z} - h^*_n e^{i \kappa^*_n z} \right)
        \end{pmatrix}^\dagger
        \begin{pmatrix}
        \frac{\omega_n}{c \kappa_n} \left( e^{i \kappa_n z} - h_n e^{-i \kappa_n z} \right) \\
        \mp \left( e^{-i \kappa_n z} + h_n e^{i \kappa_n z} \right) \\
        0
        \end{pmatrix} \\
        & = \pm \frac{2a \omega_n}{c \kappa_n} \left( h_n - h_n^{-1} \right).
    \end{split}
    \label{eq::numerator_berry_potential}
\end{equation}

\noindent Hence, the Berry potential can be explicitly written as:

\begin{equation}
    k \mathcal{A}_{\phi, n}^{\pm} =  \pm i \frac{2a \omega_n}{c \kappa_n} \left( h_n - h_n^{-1} \right) \left[ \frac{4 a \omega_n^2}{c^2 \kappa_n^2} + \frac{i k^2}{\kappa_n^3}\left(e^{i\kappa_n a} - e^{-i \kappa_n a}\right) \left( h_n + h_n^{-1}\right) \right]^{-1}.
    \label{eq::angular_berry_potential}
\end{equation}

In order to obtain the Chern numbers from Eqs. (\ref{eq::chern_numbers_main_text}) and (\ref{eq::angular_berry_potential}), it is necessary to evaluate $e^{i \kappa_n a}$ and $h_n$ in the limits $k \to 0^+$ and $k \to \infty$. Such analysis is reported in Sec. \ref{sec:important_limits}. Using Eqs. (\ref{eq::useful_limits_k_to_infty}) and (\ref{eq::useful_limits_k_to_0}) we find that:

\begin{equation}
    \lim_{k \to \infty} k \mathcal{A}_{\phi, n}^{\pm} = 0 \quad \text{and} \quad \lim_{k \to 0^+} k \mathcal{A}_{\phi, n}^{\pm} = \mp s \times (-1)^{n+1} \text{sgn}(1 - |\rho|)
    \label{eq::limits_berry_potential}
\end{equation}

\noindent for all resistivity values $\rho \in \mathbb{R} \setminus \left\{-1,1 \right\}$ for which the $n$-th band is separable from the rest of the spectrum and the Berry potential is smooth in $k \in \left]0,\infty \right[$. The sign $\pm$ specifies the up/down polarization of the pseudospinors and $s = \pm$ the square root branch. As a result,

\begin{equation}
    \mathcal{C}_{n}^\pm = \pm s \times (-1)^{n+1} \text{sgn}(1 - |\rho|).
    \label{eq::chern_number}
\end{equation}

\begin{comment}

\noindent and the spin Chern numbers $\mathcal{C}_{n}^s \equiv (\mathcal{C}_{n}^+ - \mathcal{C}_{n}^-)/2$ read

\begin{equation}
    \mathcal{C}_{ n}^s = (\pm) (-1)^{n+1} \text{sgn}(1 - |\rho|).
    \label{eq::spin_chern_numbers}
\end{equation}

\end{comment}

%\noindent Choosing the left-right Berry connection leads to the simplest possible analysis to obtain the Chern numbers in Eq.(\ref{eq::chern_number}). This is because it concerns no complex conjugated quantities under the picked inner-product. 
Despite the focus on non-conservative PPWs, the $\rho = 0$ case is also accounted for in the previous calculations. In fact, the biorthogonal system of left and right eigenstates can still be found for conservative models with $\hat{\mathcal{H}} = \hat{\mathcal{H}}^\dagger$. Evidently, for a conservative system, $\phi_n^{\text{L}} = \phi_n^{\text{R}}$, and in such a case we recover the Hermitian formulation of topological band theory.    

When the plates are interchanged in the guide ($\rho \to \rho^{-1}$), Eq. (\ref{eq::modal_equation}) is left invariant. Thus, it still determines the transverse wavenumbers $\kappa_n$ and natural frequencies $\omega_n$ in the transformed system. However, the weight $h_n$ relating the coefficients of the scalar function $h(z) = Ae^{i \kappa_n z} + Be^{-i \kappa_n z}$ as $B = h_n A$ transforms as $h_n \to h_n^{-1}$. According to Eqs. (\ref{eq::denominator_berry_potential}) and (\ref{eq::numerator_berry_potential}), the azimuthal component of the Berry potential $\mathcal{A}_{\phi,n}^\pm$ must then invert its sign in the new guide, $\mathcal{A}_{\phi,n}^\pm \to -\mathcal{A}_{\phi,n}^\pm$. This means that the spin Chern numbers (for the same frequency band) in two waveguides related by parity-symmetry have opposite signs.

\begin{comment}
In the $k \to \infty$ limit, the pseudospinors in Eq. (\ref{eq::explicit_pseudospinors}) can be chosen real-valued. Indeed, since $h_n \approx e^{i \kappa_n a}$ when $k$ is large, the corresponding electromagnetic fields in Eq. (\ref{eq::field_structure_TE}) become 

\begin{equation}
    \begin{cases}
    H_{\bm{k}, x} (z) \approx 2 A e^{i \kappa_n a/2} \cos \left[ \kappa_n(z-a/2) \right] \\
    H_{\bm{k}, z} (z) \approx -2i A e^{i \kappa_n a/2} \frac{k}{\kappa_n} \sin \left[ \kappa_n(z-a/2) \right] \\
    E_{\bm{k}, y} (z) \approx - 2i A e^{i \kappa_n a/2} \frac{\mu_0 \omega_n}{\kappa_n} \sin \left[ \kappa_n(z-a/2) \right]
    \end{cases}.
    \label{eq::field_structure_k_inf}
\end{equation}

\noindent While the transverse wavenumbers $\kappa_n$ are asymptotically real-valued and bounded, the corresponding frequencies $\omega_n$, though real-valued, diverge to infinity [see Sec. \ref{sec:important_limits}]. 
For $k \to \infty$, the $x$-spatial component of the magnetic field in Eq.(\ref{eq::field_structure_k_inf}) 
is negligible when compared to the other field components in Eq. (\ref{eq::explicit_pseudospinors}), 
\end{comment}

It is useful to analyze the asymptotic properties of the  pseudospinors $\bm{\Psi}^{\pm}_{\bm{k}, n}$ given by Eq. (\ref{eq::explicit_pseudospinors}).
%depend on the transverse wavenumber $\kappa_n$ and on the amplitude $h_n$. The second and third components are also proportional to $\omega_n$ and $k$, respectively. 
In the limit $k \to \infty$, $\kappa_n$ and $h_n$ are finite, but the frequency $\omega_n$ diverges [see Sec. \ref{sec:important_limits}]. Therefore, the projection of the pseudospinor along $\uvec{k}$ is negligible when compared to the other components:

\begin{equation}
    \bm{\Psi}^{\pm}_{\bm{k}, n} (z) = 2i A Z_0 e^{i \kappa_n a/2}
    \begin{pmatrix}
    0 \\
    - \frac{\omega_n}{c\kappa_n} \sin \left[ \kappa_n(z-a/2) \right] \\
    \pm \frac{k}{\kappa_n} \sin \left[ \kappa_n(z+a/2) \right] 
    \end{pmatrix}.
\end{equation}

\noindent Apart from a global phase, the $\bm{\Psi}^\pm$ spinor components  are all real-valued resulting in $\lim_{k \to \infty} k \mathcal{A}_{\phi, n}^\pm = 0$, in agreement with Eq. (\ref{eq::limits_berry_potential}).
%(\ref{eq::angular_berry_potential}).

\section{Particle-Hole Symmetry in Photonics}
\label{sec::particle_hole_symmetry_in_photonics}

Because the electromagnetic field is real-valued, the positive and negative frequency spectra must be balanced in photonic platforms. Specifically, if there is a mode with a time variation of the type $e^{-i \omega t}$, then it has a partner with time variation $e^{+i\omega^* t}$. Thereby, the spectrum of photonic systems is constrained by the particle-hole symmetry $\omega(\bm{k}) = - \omega^* (\bm{k})$, following the terminology commonly adopted in topological photonics. 

\begin{figure}[h]
\includegraphics[width=0.6\textwidth]{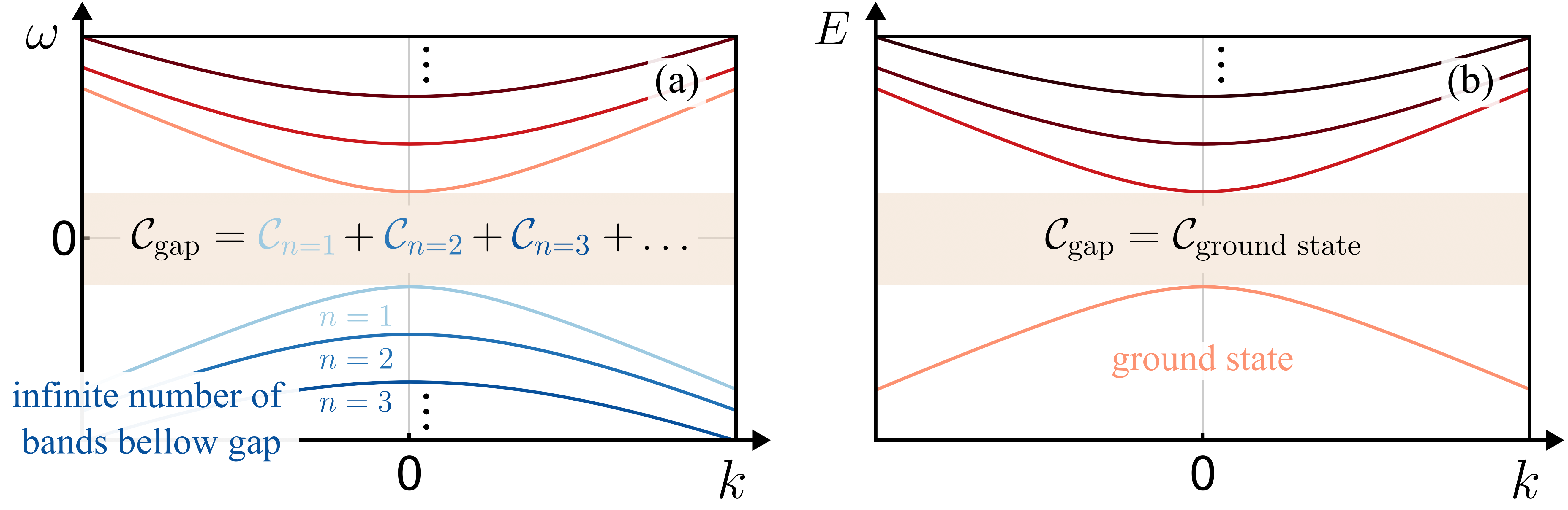}
\caption{\label{fig::ground_state}Schematic band structures of conservative (a) photonic and (b) electronic systems. Different from the electronic case, typically the photonic spectrum has an infinite number of bands bellow the gap due to the particle-hole symmetry. Hence, there is no ground state. The gap Chern number is determined by the sum of all the Chern numbers of all bands below the gap. For the photonic case, $\mathcal{C}_{\text{gap}}$ is given by an infinite sum of integers, and thus it may be ill-defined \cite{filipa_ill_defined}.}
\end{figure}

In our parallel-plate waveguide, the boundary conditions determine an infinite number of modes with positive frequencies ($\omega' > 0$) for any fixed $\bm{k}$. The particle-hole symmetry then requires the existence of an infinite number of modes with negative frequencies ($\omega' <0$) and wave vector $- \bm{k}$. The result is that our gapped systems are characterized by an infinite number of bands below the zero-frequency band gap. Thus, unlike electronic platforms, our guide has no ground state [see Fig. \ref{fig::ground_state}]. In this situation, the standard methods of topology may breakdown \cite{filipa_ill_defined}. In Sec. \ref{sec::edge_states}, we shall show that even though our system has no "ground" it is still possible to use the Chern numbers of the individual bands to predict the emergence of protected edge states.

\section{\label{sec:important_limits} Useful limits}

For convenience we decompose the transverse wavenumber and frequency into real and imaginary components: $\kappa_n = \kappa_n' + i \kappa_n''$ and $\omega_n = \omega_n' + i \omega_n''$. When $k=0$, the modal equation (\ref{eq::modal_equation}) takes the form
% vanishes for all resistivity values $-\infty < \rho < \infty$
\begin{equation}
\begin{split}
    \lim_{k \to 0^+} e^{2i \kappa_n a} & = \lim_{k \to 0^+} \frac{s \times \sqrt{k^2 + \kappa_n^2} + \rho \kappa_n}{s \times \sqrt{k^2 + \kappa_n^2} - \rho \kappa_n} \frac{ s \times \sqrt{k^2 + \kappa_n^2} \rho + \kappa_n}{s \times \sqrt{k^2 + \kappa_n^2} \rho - \kappa_n}  = - \left| \frac{s \times \rho + 1}{s \times \rho - 1} \right|^2 \\
    & = \exp\left(2i \left[ \frac{\pi}{2a}(2 n - 1) - s \times i\frac{\mathcal{F}(\rho)}{a} \right] a\right) \quad \text{with} \quad \mathcal{F}(\rho) \equiv \text{sgn}(\rho) \ln \left|\frac{|\rho| + 1}{|\rho| - 1} \right|
\end{split}
\label{eq::limit_modal_equation_k_to_0}
\end{equation}

\noindent and $n \in \mathbb{N}$. The sign $s = \pm$ selects the square root branch and is omitted for clarity ($\kappa_n^{(s=\pm)} \to \kappa_n$). Without loss of generality, we take $\kappa'_{n}>0$. Evidently, 

\begin{equation}
    \kappa_n^{(s=\pm)}(k=0) = \frac{\pi}{2a} (2n-1) - s\times i \frac{\mathcal{F}(\rho)}{a}.
    \label{eq::transverse_wavenumbers_null_k}
\end{equation}

On the other hand, in the $k \to \infty$ limit with $\rho \neq 0$, the modal equation (\ref{eq::modal_equation}) becomes

\begin{equation}
     e^{2i \kappa_m a} = 1 \quad ({k \to \infty \,\,\,\, \rm{\,with\,} \,\,\,\, \rho \neq 0}).
    \label{eq::limit_modal_equation_k_to_infty}
\end{equation}
%\noindent when $\lim_{k \to \infty, \rho \neq 0} \kappa_m /k = 0$, that is to say, when $\kappa_m$ are well-behaved. 
As a result, it follows that

\begin{equation}
    \kappa^{(s = \pm)}_m (k \to \infty; \rho \neq 0) \to \frac{\pi}{2a} 2m \quad \text{with} \quad m \in \mathbb{N}.
    \label{eq::transverse_wavenumbers_infinite_k}
\end{equation}

\noindent Numerical calculations show that
the solutions labeled by index $n$ in Eq.(\ref{eq::transverse_wavenumbers_null_k}) are analytically continued to the solutions of (\ref{eq::transverse_wavenumbers_infinite_k}) with $m = n$. Thus, we can write
\begin{equation}
    \kappa_n^{(s = \pm)}(k \to \infty; \rho \neq 0) \to \frac{\pi}{2a} 2n.    \label{eq::transverse_wavenumbers_k_to_infty_rho_not_0}
\end{equation}
Here, $\kappa_n(k)$ ($0 \leq k < \infty$) represents the $n$-th transverse wavenumber band of the waveguide.

%Label the complete branches $\kappa_n(k)$ ($0 \leq k < \infty$) with the integer $n$ that indexes the $k = 0$ solutions shown in Eq.(\ref{eq::transverse_wavenumbers_null_k}). Analytically, there seems to be no way of predicting how the $m$ integers in Eq.(\ref{eq::transverse_wavenumbers_infinite_k}) for the $k= \infty$ solutions link to each $n$. However, numerical simulations reveal that
%\begin{equation}
 %   \kappa_n^{(s = \pm)}(k \to \infty; \rho \neq 0) \to \frac{\pi}{2a} 2n \quad \text{i.e.} \quad m = n.
    %\label{eq::transverse_wavenumbers_k_to_infty_rho_not_0}
%\end{equation}

For completeness, we note that when $\rho = 0$, the modal equation (\ref{eq::modal_equation}) becomes independent of $k$, and reduces to $e^{2i\kappa_n a} = -1$. The corresponding transverse wavenumbers are $\kappa_n^{(s=\pm)}(k; \rho = 0) = \frac{\pi}{2a} (2n-1)$ with $n \in \mathbb{N}$, and match the $k=0$ solutions in Eq.(\ref{eq::transverse_wavenumbers_null_k}) with $\mathcal{F}(\rho = 0)$.  

\begin{comment}
\begin{equation}
    \lim_{\rho \to 0} e^{2i\kappa_n a} = -1,
    \label{eq::modal_equation_limit_k_to_infty_but_rho_is0}
\end{equation}

\noindent and so are the wavenumber profiles

\begin{equation}
    \kappa_n^{(s=\pm)}(k; \rho = 0) = \frac{\pi}{2a} (2n-1) \quad \text{with} \quad n \in \mathbb{N}
    \label{eq::transverse_wavenumbers_limit_k_infinity_rho_is0}
\end{equation}

\noindent that match the $k=0$ solutions in Eq.(\ref{eq::transverse_wavenumbers_null_k}) with $\mathcal{F}(\rho = 0)$.

The results in Eqs. (\ref{eq::transverse_wavenumbers_k_to_infty_rho_not_0}) and (\ref{eq::transverse_wavenumbers_limit_k_infinity_rho_is0}) indicate that well-behaved wavenumber branches are always finite when $k \to \infty$, regardless of the value of $\rho$. Consequently, the frequency bands $\omega_n^{(s=\pm)}(k)$ degenerate in the same limit as

\end{comment}

From Eq. (\ref{eq::transverse_wavenumbers_k_to_infty_rho_not_0}), it is evident that the asymptotic form of the eigenfrequencies is ruled by

\begin{equation}
    \lim_{k \to \infty} \omega_n^{(s = \pm)}(k) = \lim_{k \to \infty} s \times c k.
    \label{eq::vacuum_response}
\end{equation}

\noindent The preceding results can be used to determine $e^{i \kappa_n a}$ and $h_n = e^{ i \kappa_n a} \frac{\omega_n - c \rho \kappa_n}{\omega_n + c \rho \kappa_n}$ in the limits $k \to 0^+$ and $k \to \infty$. These limits are useful to evaluate Chern numbers in Eq. (\ref{eq::chern_number}). Straightforward calculations show that
\begin{equation}
    \lim_{k \to \infty, \rho \neq 0} e^{i \kappa_n a} = \lim_{k \to \infty, \rho \neq 0} h_n = (-1)^{n} \quad \text{and} \quad \lim_{k \to \infty, \rho = 0} e^{i \kappa_n a} = \lim_{k \to \infty, \rho = 0} h_n = i(-1)^{n+1}
    \label{eq::useful_limits_k_to_infty}
\end{equation}

\noindent and

\begin{equation}
    \lim_{k \to 0^+} e^{i \kappa_n^{(s = \pm)} a} = i(-1)^{n+1} \left| \frac{|\rho| + 1}{|\rho| - 1} \right|^{s \times \text{sgn}(\rho)} \quad \text{and} \quad \lim_{k \to 0^+} h_n = i(-1)^{n + 1} \text{sgn} (1- |\rho|).
    \label{eq::useful_limits_k_to_0}
\end{equation}

\section{\label{eq::band_structure_near_phase_transition} Band structure near the Phase transition} 

\begin{figure}[h]
\includegraphics[width=.85\textwidth]{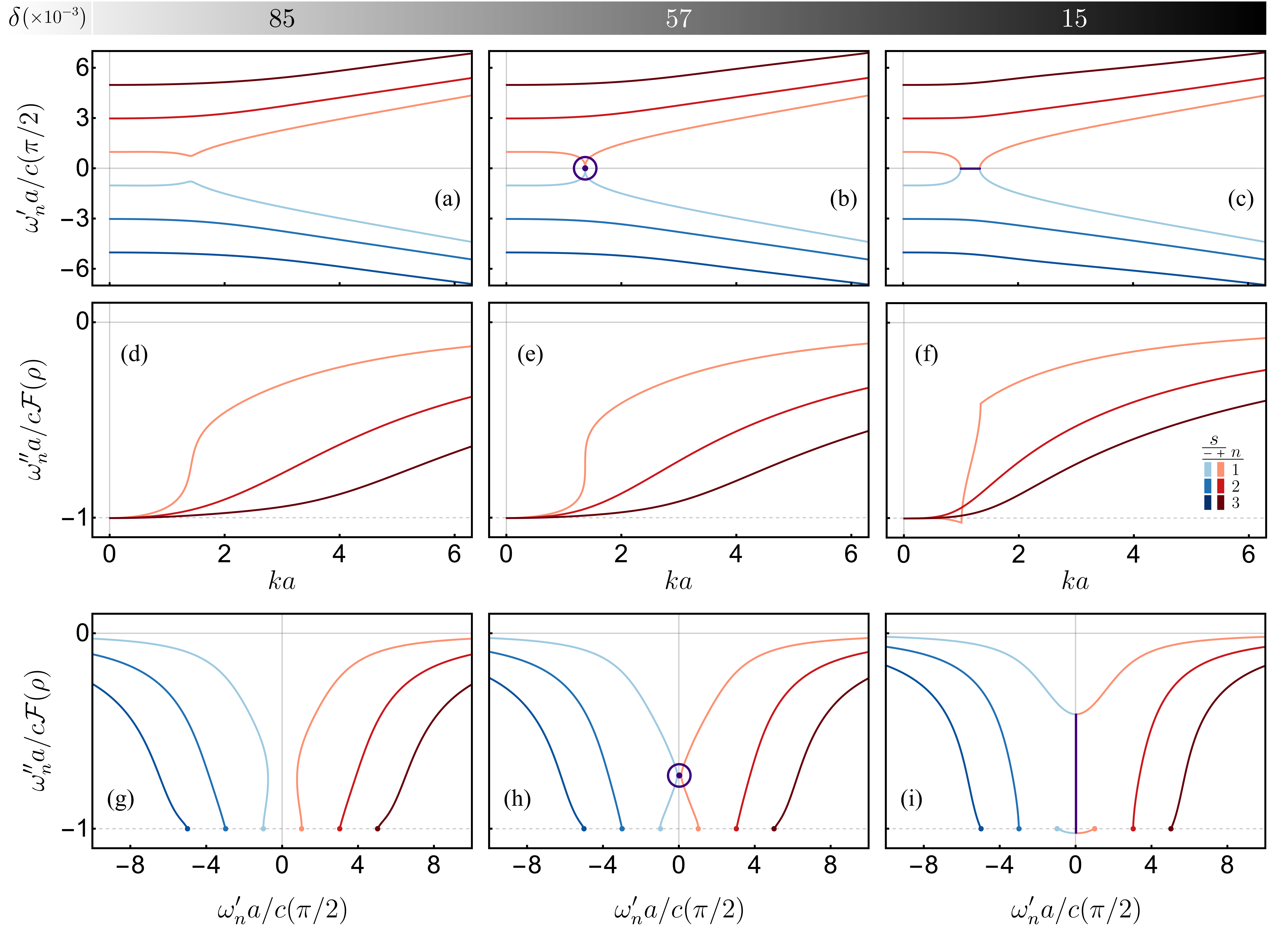}
\caption{\label{fig::band_structure_separated}(a, b, c) Real $\omega_{n}'$ and (d, e, f) imaginary $\omega_{n}''$ components of the frequency bands as functions of the in-plane momentum magnitude $k$, for $n \in \left\{ 1,2,3\right\}$ and both signs of the square root $s = \pm$. The branches $\omega_{n}^{(s=+)\prime\prime}$ and $\omega_{n}^{(s=-)\prime\prime}$ coincide in (d, e, f). (g, h, i) Locus of the band structure in the complex plane. The resistivity $\rho$ is determined by (g) $\delta = 85 \times 10^{-3}$, (h) $\delta = 57 \times 10^{-3}$ and (i) $\delta = 15 \times 10^{-3}$ with $\rho = 1-\delta$. The degeneracies between square root branches are indicated in purple.}
\end{figure}

In Figs. 2(c), 2(d) and 2(e) of the main text, the lowest-order frequency bands of propagating modes inside the PPW are projected onto the complex plane for three values of $\delta = 1 - |\rho|$ near the critical resistivity $\rho = 1$. Each curve represents the locus in the complex-plane of $\omega^{(s=\pm)}_{n=1} (k)$ for all $k$ real-valued and a fixed square root sign $s = \pm$. For a better understanding of how the frequency solutions evolve with the in-plane momentum magnitude, the real and imaginary components $\omega_{n}'$ and $\omega_{n}''$ are shown separately as functions of $k$ in Fig. \ref{fig::band_structure_separated}, not only for the $n=1$ bands, but also for the bands $n=2 \,\text{and}\, n=3$. 

In Supplementary Movie 1, we present the evolution of the band structure in the complex plane as the resistivity varies continuously from $\rho = 0.762$ to $\rho = 0.987$. The positive ($s = +$) and negative ($s = -$) bands are displayed for $n = 1,2,3$. The coalescence of the $n=1$ bands is highlighted in purple after the resistivity crosses the critical value $\rho = 0.9434$. For a better understanding of the dependence on $\rho$, the imaginary parts $\omega_n^{(s=\pm)\prime \prime}$ are not normalized to $\mathcal{F}(\rho)$ in the supplementary movie. The frequency loci at $k=0$ are represented by solid dots. 

\section{Edge States of the PEC-PMC Guide Terminated by an Opaque Wall}
%\section{Edge States Between the PEC-PMC Guide and a $\mathcal{P}\mathcal{T}\mathcal{D}$-Symmetric Wall}
\label{sec::edge_states}

%\subsection{Waveguide with a Curved Lateral Wall}

\begin{figure}[h]
\includegraphics[width=0.6\textwidth]{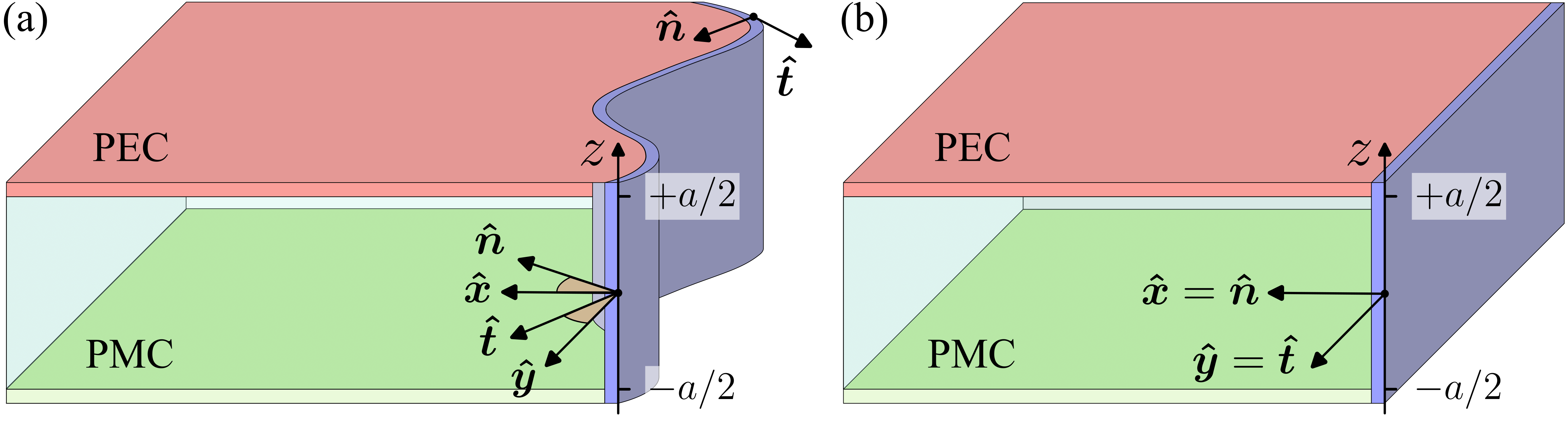}
\caption{\label{fig::boundary_ppw} Sketch of the PEC-PMC waveguide terminated with an  
%$\mathcal{P}\mathcal{T}\mathcal{D}$-symmetric
opaque lateral wall (purple). (a) Arbitrarily shaped-vertical wall. The normal and tangent unit vectors to the wall, $\uvec{n}$ and $\uvec{t}$, are functions of the position on the wall, but everywhere perpendicular to the $z$-axis. (b) Straight edge. In this case, $\uvec{n}$ and $\uvec{t}$ are constant and coincide with the unit vectors of the basis of the $xoy$-plane.}
\end{figure}

Suppose that the PPW with a PEC plate on the top and a PMC plate on the bottom is terminated by an arbitrarily shaped curved vertical wall [see Fig. \hyperref[fig::boundary_ppw]{S3(a)}]. 
%We start with a general case of a vertical and curved boundary. 
The unit vector $\uvec{n}$ normal to the wall lies on the $xoy$-plane ($\uvec{n} \cdot \uvec{z} = 0$) and is a function of the position on the wall. $\uvec{n}$ points to the interior of the guide.
\begin{comment}
written as $\uvec{n} = \cos \theta \, \uvec{x} + \sin \theta \, \uvec{y}$, with $\theta = \theta(x,y)$
\end{comment}

\subsection{Anisotropic and $\mathcal{P}\mathcal{T}\mathcal{D}$-Symmetric Curved Lateral Wall}

We consider that the wall is characterized by an anisotropic surface impedance, so that the electric and magnetic fields satisfy the Leontovich boundary condition

\begin{equation}
    \bm{Z} \cdot \left( \uvec{n} \times \bm{H}_{\text{tan}} \right) = \bm{E}_{\text{tan}},
\label{eq::leontovich_boundary_condition_lateral_wall}
\end{equation}

\noindent where $\bm{F}_{\text{tan}} = \bm{F} - \left(\bm{F} \cdot \uvec{n} \right) \uvec{n}$ is the tangential field ($\bm{F} \in \left\{ \bm{E}, \bm{H}\right\}$) and $\bm{Z} = Z_\parallel \, \uvec{t} \otimes \uvec{t} + Z_z \, \uvec{z}\otimes\uvec{z}$
%\bm{Z} = Z_\parallel \left(\uvec{x}\otimes\uvec{x} + \uvec{y}\otimes\uvec{y} \right) + Z_z \, \uvec{z}\otimes\uvec{z}$ 
is the surface impedance. Here, $\otimes$ denotes the tensor product of two vectors. %, written in matrix notation as $\uvec{x} \otimes \uvec{x} = \uvec{x} \cdot \uvec{x}^\intercal$. 
Anisotropic surface impedances can be implemented using corrugated conducting plates \cite{kildal_0, kildal_2}. In particular, surfaces such that one of the components $Z_{\parallel}$ or $Z_{z}$ vanishes and the other diverges to infinity mimic soft or hard acoustic scatterers in electromagnetics \cite{kildal_0, kildal_2}. Here, we consider the general case where the impedances are finite and dispersive.

The 
%surface impedance 
components $Z_{\parallel}$ or $Z_{z}$ are constrained by $\mathcal{P} \mathcal{D}$-symmetry. As the surface impedance of the lateral wall is independent of $z$, one sees from Eq. (\ref{eq::pd_transformation}) that the Leontovich boundary condition is compatible with the $\mathcal{P} \mathcal{D}$-symmetry of the guide only if $\bm{Z} \cdot \left( \uvec{n} \times \left[Z_0^{-1} \bm{V} \cdot \bm{E} \right]_{\text{tan}} \right) = \left[ Z_0 \bm{V} \cdot \bm{H} \right]_{\text{tan}}$.
%${\bf{Z}} \cdot \left( {{\bf{\hat n}} \times {{\left[ {\frac{1}{{{Z_0}}}{\bf{V}} \cdot {\bf{E}}} \right]}_{\tan }}} \right) = {Z_0}{\left[ {{\bf{V}} \cdot {\bf{H}}} \right]_{\tan }}$
It is straightforward to check that this condition is consistent with (\ref{eq::leontovich_boundary_condition_lateral_wall}) if and only if

\begin{equation}
    Z_{\parallel} Z_{z} = Z_0^2.
    \label{eq::ptd_contition_lateral_wall}
\end{equation}

\noindent Thus, the $\mathcal{P} \mathcal{D}$-invariance of the lateral wall requires that the surface impedance components satisfy the above constraint. In the Hermitian case, the impedances $Z_{\parallel}$ or $Z_{z}$ are pure imaginary numbers, so that both the wall and bulk region are $\mathcal{P}\mathcal{T}\mathcal{D}$-symmetric.

%Next, 
Using Eq. (\ref{eq::relation_fields_pseudospinors}), we may write the boundary condition  (\ref{eq::leontovich_boundary_condition_lateral_wall}) in terms of the pseudospinors. Interestingly, under the constraint (\ref{eq::ptd_contition_lateral_wall}), the pseudospinors satisfy a single scalar constraint on the lateral wall:

\begin{equation}
    \Psi_{z}^{\pm} (z) = \mp \left( Z_z/Z_0 \right) \Psi_{t}^{\pm} (-z).
\label{eq::lateral_wall_single_boundary_condition_pseudospinors}
\end{equation}

\noindent 
The subscripts $z$ and $t$ refer to the pseudospinor components in the right-handed orthogonal basis $\left(\uvec{n}, \uvec{t}, \uvec{z}\right)$.

\subsection{Hamiltonian Decomposition}
\label{subsec::hamiltonian_decomposition}

Next, we show that the bulk label $n$ remains a good ``quantum number'' for the modes of the electromagnetic field, even in the presence of the boundary wall. In other words, the curved wall does not mix bulk modes with different $n$ in our problem.

To demonstrate this, we look for excitations of the electromagnetic field
such that
%for which 
the dependence on $z$ is separable from the dependence on $x$ and $y$, for all the field components. This is possible because the wall is vertical ($\uvec{n} \cdot \uvec{z} = 0$) and the surface impedance is independent of the $z$-spatial coordinate. Specifically, inspired by the structure of bulk modes with index $n$ in a PEC-PMC guide, we pick the following field \textit{ansatz}

\begin{equation}
    \bm{\Psi}_{n}^{\pm} (\bm{r}, t) :=
	\begin{pmatrix}
	\sin \left[ \kappa_n (z-a/2) \right] \psi_x (x,y) \\
	\sin \left[ \kappa_n (z - a/2) \right] \psi_y (x,y) \\
	\cos \left[ \kappa_n (z - a/2) \right] \psi_z (x,y) 
	\end{pmatrix} e^{-i \omega t} \quad \text{with} \quad
	\kappa_n = (2n - 1) \frac{\pi}{2a} \,\, n \in \mathbb{N}, 
\label{eq::ansatz_pseudospinors}
\end{equation}

\noindent 
%for the $n$-th pseudospinor with $n$-th frequency that is a solution of Maxwell's equations (\ref{eq::decoupled_maxwell_equations}). 
Note that we use the values of the transverse wavenumber $\kappa_n$ found in Sec. \ref{sec:important_limits} for a conservative ($\rho = 0$) guide.
The ansatz automatically guarantees that the boundary conditions on the top/bottom walls of the waveguide are satisfied. Indeed, as the electric/magnetic fields tangential to the PEC/PMC plate must vanish we need that: $E_x (a/2) = E_y (a/2) = 0$ and $H_x(-a/2) = H_y (-a/2) = 0$. We use the shorthand notation $F(\pm a/2) \equiv F(z = \pm a/2)$. From Eq. (\ref{eq::relation_fields_pseudospinors})
%Since
%\begin{equation}
%    \begin{pmatrix}
%        \bm{E} (z) \\
%        Z_0 \bm{H} (z)
%    \end{pmatrix} = 
%    \begin{pmatrix}
%        \bm{\Psi}^{\pm} (z) \\
%        \mp \bm{V} \cdot \bm{\Psi}^{\pm} (-z)
%    \end{pmatrix},
    %\label{eq::relation_fields_pseudospinors}
%\end{equation}
%\noindent 
it follows that these boundary constraints may be written in terms of the pseudospinors simply as $\Psi_x (a/2) = \Psi_y (a/2) = 0$, which is trivially satisfied by our \textit{ansatz} because $\sin 0 = 0$. 

Feeding the pseudospinors in Eq. (\ref{eq::ansatz_pseudospinors}) into the Maxwell's equations (\ref{eq::decoupled_maxwell_equations}) leads to

\begin{equation}
\pm i
	\begin{pmatrix}
		\cos \left[ \kappa_n (z + a/2) \right] \left( - \kappa_n \psi_y + \partial_y \psi_z \right) \\ 
		\cos \left[ \kappa_n (z + a/2) \right] \left( \kappa_n \psi_x -  \partial_x \psi_z \right) \\
		\sin \left[ \kappa_n (z + a/2) \right] \left( - \partial_y \psi_x + \partial_x \psi_y \right)
	\end{pmatrix} = \frac{\omega}{c}
	\begin{pmatrix}
		\sin \left[ \kappa_n (z - a/2) \right] \psi_x \\ 
		\sin \left[ \kappa_n (z - a/2) \right] \psi_y \\
		\cos \left[ \kappa_n (z-a/2) \right] \psi_z
	\end{pmatrix}.
 \label{eq::mid_step}
\end{equation}

\noindent Since

\begin{equation}
    \cos \left[ \kappa_n (z + a/2) \right] = (-1)^{n} \sin \left[ \kappa_n(z - a/2) \right] \quad \text{and} \quad \sin \left[ \kappa_n(z + a/2) \right] = (-1)^{n+1} \cos \left[ \kappa_n (z - a/2) \right],
    \label{eq::auxiliary_equations}
\end{equation}

\noindent the functions of $z$ on each side of Eq. (\ref{eq::mid_step}) can be matched componentwise. Hence, we find

\begin{equation}
\underbrace{
	\pm i (-1)^{n}
	\begin{pmatrix}
		0 & - \kappa_n &  \partial_y \\
		\kappa_n & 0 & - \partial_x \\
		\partial_y & - \partial_x & 0 
	\end{pmatrix}
}_{\textstyle \equiv \hat{\mathcal{H}}^{\pm}_{n}}
	\bm{\psi} (x,y) = \frac{\omega}{c} \bm{\psi} (x,y) \quad \text{with} \quad \bm{\psi} (x,y) = 
	\begin{pmatrix}
		\psi_x (x,y) \\
		\psi_y (x,y) \\
		\psi_z (x,y) 
	\end{pmatrix},
\label{eq::separated_equation_maxwell}
\end{equation}

\noindent i.e., an equation for the vector $\bm{\psi}$ alone. 

The last step to demonstrate that $n$ remains a good quantum number is to show that the \textit{ansatz} is compatible with the boundary condition on the lateral wall (\ref{eq::lateral_wall_single_boundary_condition_pseudospinors}).
Feeding Eq. (\ref{eq::ansatz_pseudospinors}) into Eq. (\ref{eq::lateral_wall_single_boundary_condition_pseudospinors}) leads to

\begin{equation}
    \cos \left[ \kappa_n (z - a/2) \right]\, \uvec{z} \cdot \bm{\psi} = \pm \frac{Z_z}{Z_0} \sin \left[ \kappa_n (z + a/2) \right] \, \uvec{t} \cdot \bm{\psi}.
\end{equation}

\noindent The $z$ dependence on both sides of the above equation is compatible due to Eq. (\ref{eq::auxiliary_equations}). Therefore, the boundary constraint on the anisotropic and $\mathcal{P}\mathcal{T}\mathcal{D}$-symmetric lateral wall reduces to

\begin{equation}
    \uvec{z} \cdot \bm{\psi} = \pm i \chi_z (-1)^{n} \, \uvec{t} \cdot \bm{\psi},
    \label{eq::boundary_condition_psi_vector}
\end{equation}

\noindent with $\chi_z = i Z_z/Z_0$ the real-valued normalized reactance along the $z$-direction. This last result confirms that the lateral wall does not couple bulk modes associated with different quantum numbers $n$. 

\begin{figure}[h]
\includegraphics[width=.62\textwidth]{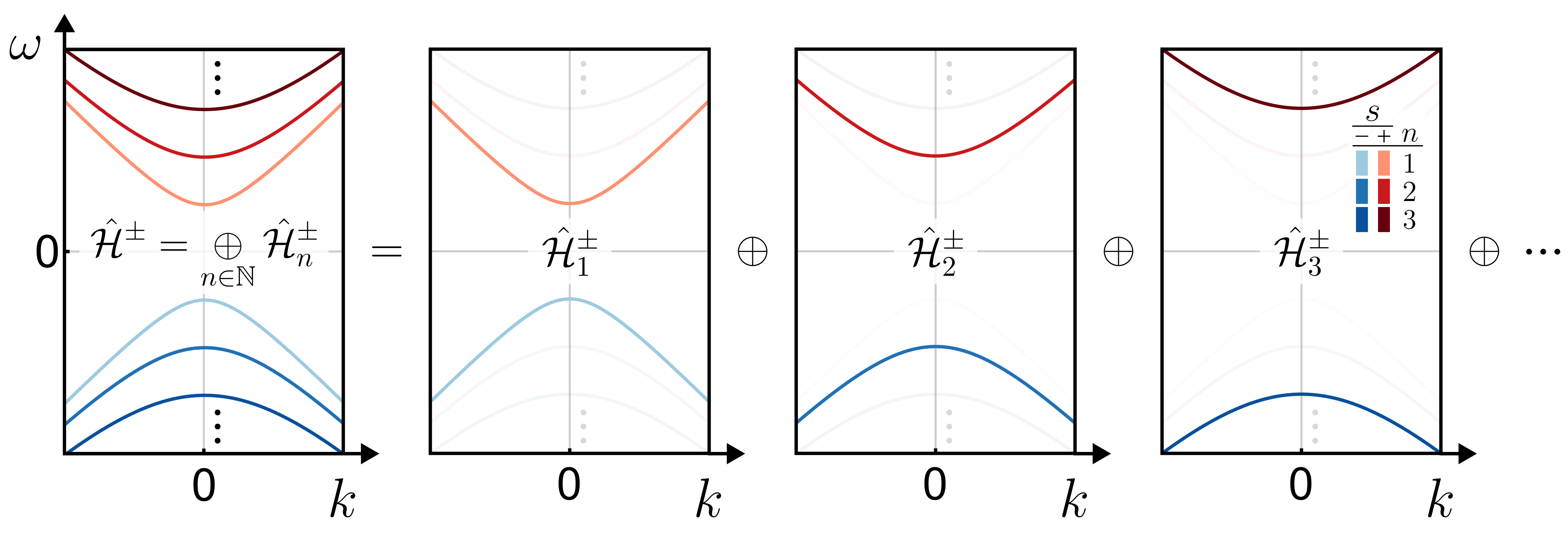}
\caption{\label{fig::hamiltonian_decomposition}Illustration of the Hamiltonian decomposition in Eq. (\ref{eq::direct_sum_hamiltonian}). The Hamiltonian of the edge waveguide $\hat{\mathcal{H}}^{\pm}$ is the direct sum of two-band Hamiltonians $\hat{\mathcal{H}}^{\pm}_{n}$ that act each only on the pseudospinors with quantum number $n$.}
\end{figure}

The previous analysis shows that the global Hamiltonian can be written as the direct sum of the Hamiltonians of the waves with quantum number $n$. This property is illustrated pictorially in Fig. \ref{fig::hamiltonian_decomposition}. The figure represents the band structure of each Hamiltonian-component in the bulk-region. However, it is crucial to note that the decomposition is not restricted to bulk modes, but it still holds true when  the waveguide is terminated by the anisotropic and $\mathcal{P} \mathcal{T} \mathcal{D}$-symmetric wall. Indeed, as previously noted, the wall
%waveguide 
does not couple pseudospinors labelled by different integers. Thereby, the eigenspaces generated by the waves with quantum number $n$ do not mix. 
Our system is effectively equivalent to an infinite set of totally decoupled waveguides, each described by an operator $\hat{\mathcal{H}}^{\pm}_n$. In particular, our analysis demonstrates that the edge states with quantum number $n$ that propagate confined to the lateral wall do not scatter into other modes (labelled by $m \neq n$), regardless of the shape of the interface. 

Let us now focus on the  Hamiltonian of the pseudospin class ``$+$". From the previous discussion, it may be decomposed as follows:
\begin{equation}
    \hat{\mathcal{H}}^{+} = \hat{\mathcal{H}}^{+}_1 \oplus \hat{\mathcal{H}}^{+}_2 \oplus \dots = \mathop{\oplus}_{n \in \mathbb{N}} \hat{\mathcal{H}}^{+}_{n},
    \label{eq::direct_sum_hamiltonian}
\end{equation}
where $\hat{\mathcal{H}}^{+}_n$ is the operator defined in Eq. (\ref{eq::separated_equation_maxwell}),
%???
which rules the waves with quantum number $n$ with the field structure shown in Eq. (\ref{eq::ansatz_pseudospinors}).
%???.
As illustrated in Fig. \ref{fig::hamiltonian_decomposition}, $\hat{\mathcal{H}}^{+}_n$ describes a two-band system. Hence, it has a single band below the gap, and its topology is well-defined. In particular, the number of edge states associated with the quantum number $n$ can be predicted using the bulk-edge correspondence. The implications of this property will be further discussed in the next subsection.

\begin{comment}
Therefore the global Hamiltonian of the waveguide with the $\mathcal{P}\mathcal{T}\mathcal{D}$-lateral wall can be written as direct sum of the Hamiltonians of the waves with quantum number $n$: $\hat H = \hat H_1^ \pm  \oplus \hat H_2^ \pm  \oplus ... \oplus \hat H_n^ \pm ...$ As illustrated in Fig. ???, each of these Hamiltonians describes a two-band model. In particular, the Hamiltonian $\hat H_n^ \pm $ has a single band  below the gap. Thus, its topology is well-defined and the number of edge states associated with the quantum number $n$ can be predicted using the bulk-edge correspondence. Note that our analysis demonstrates that the edge states with quantum number $n$ that propagate confined to the lateral wall do not not scatter into other modes (labelled by $m \neq n$), for an arbitrarily shaped interface.
\end{comment}

For completeness, it is interesting to note that 
the system of differential equations (\ref{eq::separated_equation_maxwell}) is equivalent to solving the Helmholtz equation

\begin{equation}
    - \left( \partial_x^2 + \partial_y^2 \right) \psi_z = \left( \frac{\omega^2}{c^2} - \kappa_n^2 \right) \psi_z
    \label{eq::helmholtz_equation_psi_z}
\end{equation}

\noindent for the $z$-component of $\bm{\psi}$, with the remaining 
%pseudospinor 
ones given by

\begin{equation}
    \begin{cases}
		   \psi_x = \left(\frac{\omega}{c} - \frac{ \kappa_n^2  c}{\omega}  \right)^{-1} \left(- \frac{\kappa_n c}{\omega} \partial_x  \pm i (-1)^{n} \partial_y \right) \psi_z \\
		 \psi_y = \left(\frac{\omega}{c} - \frac{ \kappa_n^2  c}{\omega}  \right)^{-1} \left( - \frac{\kappa_n c}{\omega} \partial_y \mp i (-1)^{n} \partial_x \right) \psi_z 
	\end{cases}.
    \label{eq::remainig_components_psi_x_and_psi_y}
\end{equation}

\noindent Thus, we have reduced the task of solving Maxwell's equations inside the PPW to a 2-dimensional problem. 
%Indeed, Eqs. (\ref{eq::helmholtz_equation_psi_z}) and (\ref{eq::remainig_components_psi_x_and_psi_y}) determine the electrodynamics on the $xoy$-plane for the pseudospinor solutions in Eq. (\ref{eq::ansatz_pseudospinors}) that forcibly satisfy the boundary conditions on the conductor plates of the waveguide and also account for variation along the $z$-coordinate. 
 
\subsection{Edge States for a Straight Lateral Wall}

For the case of a straight edge, we can deduce the dispersion relation of edge states analytically. We suppose that the lateral wall is defined by $x=0$, so that the edge mode propagates along the $y$-axis as illustrated in Fig. \hyperref[fig::boundary_ppw]{S3(b)}. For this setup, the tangential unit vector is constant along the wall, $\uvec{t} = \uvec{y}$. We take 

\begin{equation}
    \psi_z = e^{i \bm{k} \cdot \bm{r}},
    \label{eq::psi_z_choice}
\end{equation}

\noindent with $\bm{k} = k_x \, \uvec{x} + k_y \, \uvec{y}$ the in-plane wave vector. The pseudospinor component $\psi_z$  is a solution of the Helmholtz equation (\ref{eq::helmholtz_equation_psi_z}) with $\omega_n^2 / c^2 = \kappa_n^2 + k^2$ ($k^2 \equiv k_x^2 + k_y^2$). The $x$ and $y$ components of the vector $\bm{\psi}$ follow from Eq. (\ref{eq::remainig_components_psi_x_and_psi_y}) and read

\begin{equation}
    \begin{cases}
        \psi_x = \left( - i \frac{\kappa_n}{k^2} k_x \mp (-1)^{n} \frac{\omega_n}{c k^2} k_y \right) e^{i \bm{k} \cdot \bm{r}} \\
        \psi_y = \left( - i \frac{\kappa_n}{k^2} k_y \pm (-1)^{n} \frac{\omega_n}{c k^2} k_x \right) e^{i \bm{k} \cdot \bm{r}}
    \end{cases}.
    \label{eq::remainig_components_choice}
\end{equation}

\noindent Note that $\psi_{x,y,z} \propto e^{i \bm{k} \cdot \bm{r}}$, so that the pseudospinors in Eq. (\ref{eq::ansatz_pseudospinors}) are ruled by the propagation factor $\bm{\Psi}_{n}^{\pm} \propto e^{i \bm{k} \cdot \bm{r}}$, similar to the bulk modes of the main text. We feed the components $\psi_z$ and $\psi_y$ in Eqs. (\ref{eq::psi_z_choice}) and (\ref{eq::remainig_components_choice}) to the boundary constraint in Eq. (\ref{eq::boundary_condition_psi_vector}) with $\uvec{t} = \uvec{y}$ to find that 

\begin{equation}
    \frac{k^2}{\kappa_n} = \chi_z \left( \pm (-1)^{n} k_y + i \frac{\omega_n}{c \kappa_n} k_x \right)
    \label{eq::pre_dispersion_equation}.
\end{equation}

We are interested in edge waves that propagate attached to the lateral wall. Therefore, it is assumed that $k_y$ and $\alpha_x \equiv - i k_x = \sqrt{k_y^2 + \kappa_n^2 - \omega_n^2/c^2}$ are real-valued propagation and attenuation constants, so that the pseudospinors vary on the $xoy$-plane as $\bm{\Psi}_{n}^{\pm} \sim e^{i k_y y} e^{- \alpha_x x}$. In these conditions, the dispersion of the edge modes with quantum number $n$ [Eq. (\ref{eq::pre_dispersion_equation})] reduces to:

\begin{equation}
    \frac{\omega_n^2/c^2 - \kappa_n^2}{\kappa_n} = \chi_z \left( \pm (-1)^{n} k_y  - \frac{\omega_n}{c \kappa_n} \sqrt{k_y^2 + \kappa_n^2 - \omega_n^2/c^2} \right).
    \label{eq::dispersion_relation_edge_modes}
\end{equation}

%\noindent The above equation determines the dispersion of the edge modes with quantum number $n$. 

%In agreement with the bulk-edge correspondence, the dispersion equation (\ref{eq::dispersion_relation_edge_modes}) predicts that for each pseudospin and each quantum number $n$ there is a single gapless unidirectional edge mode propagating in the gap of the bulk. The direction of propagation is flipped for the Hamiltonians $\hat{\mathcal{H}}_{n}^{+}$ and $\hat{\mathcal{H}}_{n}^{-}$. 

\begin{figure}[h]
\includegraphics[width=0.65\textwidth]{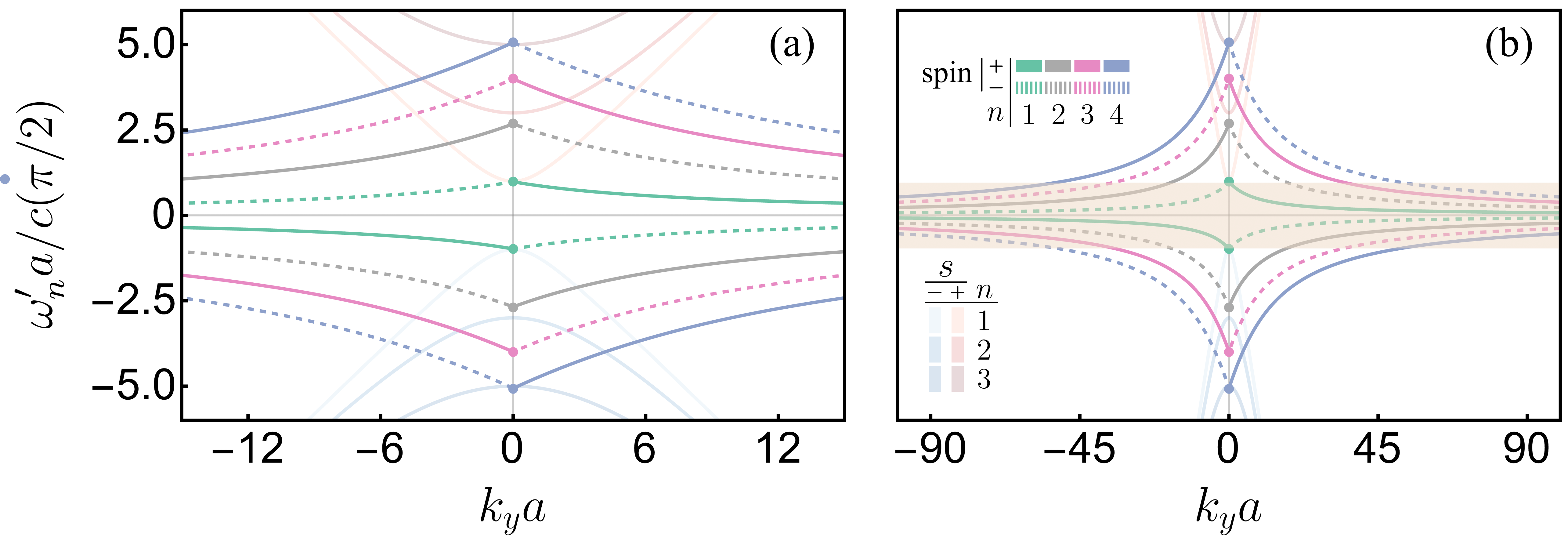}
\caption{\label{fig::dispersion_edge_states}(a) Frequency bands $\omega_n$ of bulk ($n<4$) and edge ($n<5$) waves as a function of the propagation constant $k_y$, for a normalized reactance $\chi_z = 0.12 \, \omega a/c$. Red/Blue translucent
%(Green/Purple) 
tones refer to the positive/negative 
%square root branch for 
bulk states. Other colors concern the edge modes: green for $n=1$, orange for $n=2$, purple for $n=3$ and pink for $n=4$. Edge waves with pseudospin ``$+$"/``$-$" are represented by solid/dashed bands. For a fixed pseudospin, the direction of propagation of the edge waves alternates as the quantum number $n$ changes. For the same number $n$, the edge waves have opposite propagation directions for opposite pseudospins. (b) Range of $k_y$ is extended. All edge states span the entire photonic band gap indicated by the beige rectangular region.}
%Darker colors correspond to higher-order modes. (b) and (c) represent the same as (a) but over an extended range of $k_y$. The bulk band gap is highlighted in beige.}
\end{figure}

 To illustrate the discussion, we show in Fig. \ref{fig::dispersion_edge_states} the frequency bands $\omega_n$ of the edge waves as a function of the propagation constant $k_y$, for $n<5$.
%consider the pseudospin up, i.e., choose the sign ``$+$" in the dispersion equation (\ref{eq::dispersion_relation_edge_modes}). 
%As anticipated in the above discussion, 
%We show this in Fig. \ref{fig::dispersion_edge_states} by plotting the 
%positive ($s = +$) and negative ($s = -$) 
%requency bands $\omega_n$ of the edge waves with $n<5$ as a function of the propagation constant $k_y$. 
We take $\chi_z = 0.12 \, \omega a/c$, which corresponds to an inductive $Z_z$ and a capacitive $Z_{\parallel}$. It is clear that the edged waveguide supports an \textit{infinite} number of 
%gapless 
edge waves that propagate alternately to the $+y$ and $-y$ directions, depending on the quantum number $n$. Besides, the edge waves associated with the Hamiltonians $\hat{\mathcal{H}}_{n}^{+}$ and $\hat{\mathcal{H}}_{n}^{-}$ propagate in opposite directions. A final key observation is that the edge modes are gapless: $\lim_{k_y \to \pm \infty} \omega_n = 0$. 

The properties of the edge transport discussed above and illustrated by Fig. \ref{fig::dispersion_edge_states} are in precise agreement with the topological character of the waveguide. As already discussed in Sec. \ref{subsec::hamiltonian_decomposition}, the Hamiltonian $\hat{\mathcal{H}}^{+}$ is a direct sum of partial two-band Hamiltonians $\hat{\mathcal{H}}^{+}_{n}$ that act on the pseudospinors with quantum number $n$ [see Eq. (\ref{eq::direct_sum_hamiltonian})]. Each Hamiltonian $\hat{\mathcal{H}}_{n}^{+}$ has a single (negative) frequency band  below the gap [see Fig. \ref{fig::hamiltonian_decomposition}]. Hence, its gap Chern number is well defined and given by the Chern number of that band: $\mathcal{C}_{n}^{+} = (-1)^{n}$, as discussed in the main text. The gap Chern number of the pseudospin class "+" is given by the non-convergent series $\mathcal{C}_{\text{gap}}^{+} = -1 + 1 -1 \dots$, that results from summing the Chern numbers of all the negative frequency bands. Therefore, we conclude that the Hamiltonian decomposition in Eq. (\ref{eq::direct_sum_hamiltonian}) provides a precise meaning to this infinite sum. It corresponds to the sum of the gap Chern numbers of the partial Hamiltonians $\hat{\mathcal{H}}_{n}^{+}$. In agreement with the bulk-edge correspondence, the dispersion equation (\ref{eq::dispersion_relation_edge_modes}) predicts that for each pseudospin and each quantum number $n$ there is a single gapless unidirectional edge mode propagating in the gap of the corresponding partial Hamiltonian.
It underlined here that, different from the global system, the operator $\hat{\mathcal{H}}_{n}^{+}$ has a well-defined topology, and hence its edge states are ruled by the bulk-edge correspondence.
The $n$-th term in the series $\mathcal{C}_{\text{gap}}^{+}$ determines the edge state with quantum number $n$. 

The direction of propagation of an edge state of $\hat{\mathcal{H}}_{n}^{+}$ is locked to the sign of the respective gap Chern number \cite{silveirinha_bulk_edge_proof}, $\mathcal{C}_{n}^{+} = (-1)^{n}$, in agreement with the numerical results.
%of the component $\hat{\mathcal{H}}_{n}^{+}$. 
%This connection is particularly relevant because $n$ remains a good quantum number in the presence of the wall. In particular, as it is impossible for 
As an edge state with number $n$ cannot scatter into waves with different $n$ when the shape of the lateral boundary is deformed, the edge modes are topologically protected.

%Furthermore, as $n$ remains a good quantum number in the presence of the lateral wall, it is feasible in our system to apply the bulk-edge correspondence to the global system and attribute a precise meaning to each of the terms of the divergent gap Chern number series.  .

%Indeed, the gap Chern number is the sum of the Chern numbers $\mathcal{C}^{+}_{n} = (-1)^{n}$ of the infinitely many bands that lie below the gap [see Sec. \ref{sec::particle_hole_symmetry_in_photonics}]. The direction of propagation of the $n$-th edge state is locked to the sign of the topological index $\mathcal{C}_{n}^{+} = (-1)^n$ of the $n$-th bulk mode. The total number of edge states links to the infinite number of terms in the series $\mathcal{C}_{\text{gap}}^{+}$. Herewith, we show a bulk-edge correspondence for a gap Chern number that is ill-defined.

%Note that from 
According to Eq. (\ref{eq::dispersion_relation_edge_modes}), for any $n \in \mathbb{N}$, flipping the pseudospin is equivalent to flip the propagation direction of the edge wave along the $y$-axis. This is expected from electromagnetic reciprocity. It is also consistent with the fact that the  Chern numbers associated with the "+" and "-" pseudospins differ from a minus sign: $\mathcal{C}_{n}^{-} = -\mathcal{C}_{n}^{+}$. As already mentioned, the direction of propagation of an edge wave is locked to the sign of the gap Chern number. It is also curious note that the edge modes are backward waves, and that the positive and negative frequency branches are joined at $k_y=\infty$, guaranteeing in this manner that the edge waves are gapless.

It is useful to note that Eqs. (\ref{eq::psi_z_choice}) and (\ref{eq::remainig_components_choice}) give 

\begin{equation}
    \frac{k^2}{\kappa_n} \bm{\psi} = \left( -i \, \bm{k} \pm \frac{\omega_n}{c \kappa_n} (-1)^{n} \, \uvec{z} \times \bm{k} + \frac{k^2}{\kappa_n} \, \uvec{z} \right) e^{i \bm{k} \cdot \bm{r}}.
    \label{eq::psi_vector_edge}
\end{equation}

\noindent Hence, the full pseudospinors that represent edge waves propagating along the straight lateral wall are

\begin{equation}
 \bm{\Psi}_{n}^{\pm} (\bm{r}, t) \propto \left( \sin \left[ \kappa_n (z - a/2) \right]  \left( - i \, \bm{k} \pm \frac{\omega_n}{c \kappa_n} (-1)^{n} \, \uvec{z} \times \bm{k} \right) + \frac{k^2}{\kappa_n} \cos \left[ \kappa_n (z - a/2) \right] \, \uvec{z} \right) e^{- i \omega_n t} e^{i k_y y} e^{-\alpha_x x}.
\label{eq::pseudospinors_conservative_case} 
\end{equation}

\noindent 
Curiously, the above field structure is coincident with that of a bulk mode with index $n$  but with a complex-valued wave-vector. Note that for a conservative waveguide ($\rho = 0$) one has $h_n = -e^{-i \kappa_n a}$  [see Eq. (\ref{eq::boundary_conditions_AB_result})].

\bibliography{sm}% Produces the bibliography via BibTeX.